\documentclass[10pt,letterpaper]{article}
\title{PLOS Template}
\usepackage[top=0.85in,left=2.75in,footskip=0.75in]{geometry}
\usepackage{rotating}
\usepackage{siunitx,multirow}

\usepackage{color}

\usepackage{amsmath,amssymb}
\usepackage{amstext}

\usepackage{changepage}

\usepackage{longtable}

\usepackage[utf8x]{inputenc}

\usepackage{textcomp,marvosym}

\usepackage{cite}

\usepackage{nameref,hyperref}

\usepackage[right]{lineno}

\usepackage{microtype}
\DisableLigatures[f]{encoding = *, family = * }

\usepackage[table]{xcolor}

\usepackage{array}

\newcolumntype{+}{!{\vrule width 2pt}}
\newcolumntype{R}{>{$}r<{$}} 
\newcolumntype{L}{>{$}l<{$}} 

\newlength\savedwidth

\raggedright
\usepackage[aboveskip=1pt,labelfont=bf,labelsep=period,justification=raggedright,singlelinecheck=off]{caption}

\makeatletter
\renewcommand{\@biblabel}[1]{\quad#1.}
\makeatother

\date{}

\usepackage{lastpage,fancyhdr,graphicx}
\usepackage{epstopdf,float}
\fancyheadoffset[L]{2.25in}
\fancyfootoffset[L]{2.25in}
\begin{document}
\vspace*{0.2in}

\begin{flushleft}
{\Large
\textbf\newline{A Minimal Power Model for Human Running   Performance} 
}
\newline

Matthew Mulligan\textsuperscript{1},
Guillaume Adam\textsuperscript{2},
Thorsten Emig\textsuperscript{2,3,*}
\\
\bigskip
\textbf{1} Claremont McKenna College, W.M. Keck Science Department,
Claremont, California 91711, USA
\\
\textbf{2} Massachusetts Institute of Technology, MultiScale Materials Science
for Energy and Environment, Joint MIT-CNRS Laboratory (UMI 3466),
Cambridge, Massachusetts 02139, USA
\\
\textbf{3} Laboratoire de Physique
Th\'eorique et Mod\`eles Statistiques, CNRS UMR 8626, B\^at.~100,
Universit\'e Paris-Saclay, 91405 Orsay cedex, France
\\
\bigskip

* emig@mit.edu

\end{flushleft}

\section*{Abstract}
Models for human running performances of
various complexities and underlying principles have been proposed,
often combining data from world record performances
and bio-energetic facts of human physiology. Here we present a novel,
minimal and universal model for human running performance that employs a relative metabolic power scale. The main component is a self-consistency relation for the time dependent maximal power output.  The analytic
approach presented here is the first to derive the observed
logarithmic scaling between world (and other) record running speeds
and times from basic principles of metabolic power supply. Various female and male record
performances (world, national) and also personal best performances of
individual runners for distances from 800m  to the marathon are
excellently described by this model, with mean errors of
(often much) less than $1\%$.   The model defines endurance in a way that demonstrates symmetry between long and short racing events that are separated by a characteristic time scale comparable to the time over which a runner can sustain maximal oxygen uptake. As an application of our model,
we derive personalized characteristic race speeds for different durations and distances.

\section*{Introduction}

Scientists have been fascinated by trying to explain running performance and to predict its limitations for more than 100 years.
A purely descriptive approach was employed by Kennelly as early as 1906 for speeds in racing events of animals and humans.
  For men
running events from 20 yards up to a few hundred miles he found a
power law relation between distance $d$ and duration $T$ with
$T \sim d^{9/8}$ with a relative large error of up to 9$\%$ for
distances from 100m to 50 miles (and larger errors for shorter and
longer distances) \cite{Kennely:1906aa}. 
  
Almost a century ago, in 1925 noted mathematician and physiologist A.V.~Hill proposed a power model based on metabolic energy considerations to describe the maximal power output $P_{max}(T)$ over a given duration $T$ by a hyperbolic function $P_{max}(T) = P_0+ P_1/T$ with constants $P_0$ and $P_1$ (known as the ``running curve'') \cite{Hill:1925pd}. Ward-Smith introduced a model, based on the
first law of thermodynamics, to describe performances at Olympic Games
from 1960 to 1976 with an average absolute error for the predicted
times of $0.86\%$ for distances from 100m to 10,000m
\cite{WARDSMITH:1985or}. In 1973 the mathematician Keller formulated a purely mechanical model that is based on the runner's equation of motion with a damping term \cite{Keller:1973jw}. The propulsive force is connected to the mechanical power utilized for running which is different from the overall metabolic power requirement. In analogy to purely mechanical problems, Keller assumed that the damping is linear in velocity and that the damping coefficient is constant over time. The justification for these assumptions is not validated  given that a comparison of his model to world track records from 50yards to 10,000m yields a relative large errors of about $3\%$ for distances larger than 5000m. Furthermore, both Hill's and Keller's models predict the existence of a
maximal speed that can be sustained for an infinite duration, which is
not possible from a physiological point of view and incompatible with
data on running records.

In fact, existing models appear to be
unable to {\it explain} an important observation that has 
been made already by Hill in the context of his above mentioned model: The average fractional utilization of maximal power (or the average running speed) of world record performances scales
linearly with the {\it logarithm} of the duration of the performance
\cite{Hill:1925pd}.  An interesting model that interpolates between
fundamental knowledge of human bioenergetics during exercise and
actual world record running performance was proposed by Peronnet and
Thibault \cite{Peronnet:1989dp,Peronnet:1991ix}.  Their model combines
characteristics of energy metabolism, based on Hill's hyperbolic
``running curve'' and the dynamics of oxygen uptake. However, the
fractional utilization of maximal power over a given duration is
described in their model by a phenomenological logarithmic term that
is based on observations in world records. The latter term
accounts for endurance limited sustainability of maximal aerobic power. Currently, this model is most effective in reproducing world record running performances. However, it uses a number
of fixed parameters that are assumed to be equal for all world record
performances although they have been achieved by different
athletes. In fact, many parameters can be different among individuals.
For example, running economy, i.e., the energy cost of running at a
given velocity, shows substantial inter-individual variation
\cite{Fletcher:2009uf}. These variations are observed even among well
trained elite runners.  Another quantity that is modeled as a
constant in Peronnet's and Thibault's model is the duration over which
maximal aerobic power (or VO$_{2 max}$) can be maintained during
running which they assumed to be 7 minutes. However, direct measurements of oxygen uptake have demonstrated variations of the order of one to two minutes among individuals \cite{ISI:A1994MV68700018,Bosquet:2002aa}.
From a fundamental perspective it is desirable to derive a model from basic principles of metabolic power generation and utilization that predicts  human performances without additional phenomenological input. This is the objective of the present work.

For the development of our model it is instructive to review some 
facts and experimental observations from exercise physiology.  
When developing a model that can describe running performances as obtained in world records up to the marathon distance one should realize at what relative intensities these races are performed.  All Olympic endurance events require intensities above $85\%$ 
of VO$_{2max}$ which
corresponds to the effort reached approximately in the marathon
\cite{Joyner:2008sf}. When looking at record performances, we can also assume that runner has followed an optimal carbohydrate loading strategy so that the stored amount of glycogen is permitting best possible performance. This is of importance for the half marathon and in particular the marathon distance which is raced predominantly on carbohydrate fuel with an average  respiratory gas exchange ratio of close to one for faster runners \cite{OBrien:1993aa}.

An important physiological observation is that the total energy cost of running increases linearly with the covered distance with no or a very small dependence on the running velocity
\cite{Margaria:1963zi,Leger:1984aa}. 
Hence the power output changes linearly with speed, with the slope quantifying  running economy. It is known that this running economy can vary about $30$--$40\%$ among individuals \cite{Joyner:2008sf}.  An important observation that is essential for the construction of our model is that running economy usually becomes worse with the duration of a running event.  The magnitude of the change in economy increases with duration and intensity. The
actual change is probably subject dependent and also influenced by
external conditions. We shall see below that this is an important factor in determining race velocities and endurance.  This drift in running economy has been quantified in treadmill studies with a change of $4.4\%$ for 40min at $80\%$ VO$_{2max}$, a change of $6.6\%$
for 60min at $70\%$ VO$_{2max}$, and a change of $9.5\%$ for 60min at $80\%$ VO$_{2max}$ \cite{Sproule:1998yi}.  An other study found for 60min treadmill running near $80\%$ of VO$_{2max}$ a shift of about $3\%$ in oxygen uptake \cite{Hunter:2007ph}.  Changes in running economy have been also observed during a 5km run at a constant pace eliciting about $80-85\%$ of VO$_{2max}$ with an average increase in oxygen uptake of $3.3\%$ for men and $2.0\%$ for women \cite{Thomas:1999fz}.  The reason for the increase in oxygen uptake and reduction in running economy is unknown. A number of mechanisms have been postulated in the literature but most of them are speculative \cite{OBrien:1993aa,Beis:2012xc,Grimby:1962df,Dion:2013le}. Without discussing here the various attempts that have been made for explaining this observation, we just conclude that every activated
physiological system increases its own particular energy consumption with the duration of exercise.

\section*{Methods}

\subsection*{A minimal model for running performance}

In view of the current status of theoretical descriptions of human
running performances, it appears useful to construct a {\it
  minimal and universal} model for human running performance that fulfills the following two requirements:

\begin{enumerate}

\item Based on basic  concepts and observations on metabolic power generation and utilization during running

\item Minimal number of physiological parameters that are not fixed a priori 
  
\end{enumerate}

In order to eliminate irrelevant normalization parameters from the
model (that would depend on the choice of units for energy, power,
etc.), we express our model in terms of relative quantities. We shall base the model on expedited power measured as oxygen uptake per time since this quantity can be measured directly under real conditions by mobile spirometry.  This implies a slight time dependence of oxygen uptake during prolonged exercise, even when the power output is constant, due to a change of the respiratory quotient with substrate utilization \cite{Bosch:ao}.  Also, since body weight usually changes  during prolonged exercise, we measure power or oxygen uptake always per body weight. 

While the basal metabolic rate $P_b$ is close to $1.2$W/kg\cite{Peronnet:1989dp}, its actual value is not required in the following. In fact, in the parameterization of running economy to be employed below, we chose to associate $P_b$ with the power that is obtained by linearly extrapolating the running economy to zero velocity. Hence we neglect the non-linear dependence of the energy cost on sub-running (walking)
velocities which causes no problem since our model uses the energy
cost of motion only in the linear running regime. In our model there exists a crossover power $P_m$ that we expect to be close to the maximal aerobic power associated with  maximal oxygen uptake
VO$_{2max}$ which is typically in the range of $75$ to $85$ml/(kg min) for elite runners \cite{Peronnet:1989dp}. The power $P_m$ should not be confused with the critical or the maximal power that occurs in the 3-parameter critical power model of Morton \cite{Morton:1996aa}.

We measure power relative to the base value $P_b$, in units of the
aerobic power reserve $P_m-P_b$ that is available to the runner, hence
defining the relative running power (or intensity) as
\begin{equation}
  \label{eq:6}
  p = \frac{P-P_b}{P_m-P_b}
\end{equation}
for a given power $P$ so that $0 \le p  \le 1$ for running intensities that do not require more power than provided aerobically by maximal oxygen uptake.

Following the definition of the relative running power above, we
parameterize the {\it nominal} power expenditure that is required to run at a velocity $v$, i.e., the running economy, as
\begin{equation}
  \label{eq:economy}
  p(v) = \frac{P(v)-P_b}{P_m-P_b} = \frac{v}{v_m}
\end{equation}
where $v_m$ is a crossover velocity which is the smallest velocity that elicits the nominal power $P_m$. We expect this velocity to be close to the velocity that permits the runner to spent the longest time at maximal aerobic power \cite{Billat:2000aa}.
  Here ``nominal'' implies that this power is measured for short duration and idealized laboratory conditions under which running economy is linear in velocity, at least to a very good approximation\cite{morgan1989factors}.  For velocities  $v>v_m$ the energy cost of running cannot be determined from oxygen uptake measurements due to anaerobic involvement, and the actual (non-nominal) energy cost might increase in a non-linear fashion \cite{OBrien:1993aa}. We shall see below that our model allows us to estimate this non-linear correction from the supplemental power required to race at a given velocity.

To model running performance, we need information on the maximal
duration over which a runner can sustain a given power, and hence a
certain running velocity. To quantify this information, we define
$P_{max}(T)$ as the maximal {\it average} power that can be sustained
over a duration $T$. This is the power (measured as oxygen uptake)
that is {\it nominally} required to run at a given velocity.  Hence
$P_{max}(T)$ can be used to deduce the {\it mean} running velocity of
an event of duration $T$.  In addition, we define the
instantaneous power $P_T(t)$ that a runner utilizes during a race
(defined as an event in which a fixed distance is covered in minimal
time) of duration $T$ at time $t$ with $0\le t \le T$.  $P_T(t)$
should be regarded as ``typical'' power output at time $t$ of an event
of duration $T$, meaning that a given individual runner generates a
power that in general fluctuates in time around $P_T(t)$.  It is
important to note that the instantaneous power $P_T(t)$ exceeds
$P_{max}(T)$ due to an upward shift in the required power beyond the
nominal power (for example due to decreased running economy, non-linear corrections for velocities above $v_m$). The additional energy that is required to allow for
this upward shift is assumed to grow linearly in time, providing an
 {\it supplemental power} $P_{sup}$. We expect that this power is provided
by different anaerobic and aerobic energy systems, involving different time scales
over which they mainly contribute to $P_{sup}$. Hence, we introduce a
crossover time $t_c$ that separates long (l) and short (s) running events, suggesting the parameterization
\begin{equation}
  \label{eq:2}
  P_{sup}(T) = P_{s} \quad \text{ for }  T\le t_c\, , \quad\quad
  P_{sup}(T) = P_{s} \frac{t_c}{T} + P_l \frac{T-t_c}{T}
  \quad \text{ for } T > t_c \, ,
\end{equation}
which describes the fractional contribution of energy systems during short and long events of total duration $T$. While the sharp crossover
between these regimes is an oversimplification of reality, we shall
see below that it leads to reasonable estimates. We have assumed that
there is only one crossover  time scale since there exists 
only one distinct power scale $P_m$ which is presumably set by the maximal aerobic
power. Hence we associate $t_c$ with the time scale over which maximal aerobic power can be sustained.

To construct our model, we start from the following self-consistency  relation 
\begin{equation}
  \label{eq:3}
  P_{max}(T)  + P_{sup}(T)  = \frac{1}{T} \int_{0}^{T} P_T(t) dt \, ,
\end{equation}
which states that the sum of the nominal average power and the additional
supplemental power $P_{sup}$ equals the time average of the instantaneously utilized power. 
We make the
important conjecture that the instantaneous power utilized at time $t$ equals the maximal power that can be sustained for the remaining time $T-t$ of the event \cite{Billat:2009aa}, i.e.,
\begin{equation}
P_T(t) = P_{max}(T-t) \, .
\end{equation}
Note that this implies
that the power output during a race is not constant over time but
increases towards the end of the event. 
When this relation is substituted into the self-consistency
Eq.~\eqref{eq:3}, one obtains an integral equation that determines
$P_{max}(T)$. If there would be no supplemental power ($P_{sup}=0$) then the integral equation has a constant $P_{max}(T)$ as solution since  $P_{max}(T)$ must be a non-increasing function of $T$. However, a constant solution is not acceptable since a given power cannot be sustained for all durations $T$, and hence $P_{sup}$ must be non-zero. The general solution is (for details see Appendix S1)
\begin{equation}
  \label{eq:5}
  P_{max}(T) = \left\{
    \begin{array}{ll}
    P_m - P_{s} \log  \frac{T}{t_c} 
  & \text{ for } T \le t_c \\[1em]
 P_m - P_{l} \log  \frac{T}{t_c} 
      & \text{ for } T > t_c
      \end{array}
 \right. \, ,
\end{equation}
where $P_m=P_{max}(t_c)$ is the crossover power reached at the crossover time $t_c$. We note that $P_{max}(T)$ can be compared to experimental studies of oxygen consumption during running for short durations below $t_c$, see Appendix S2.

It turns out to be useful to measure $P_{s}$ and $P_l$ as fractions of the aerobic power reserve $P_m-P_b$ by introducing two
corresponding dimensionless factors $\gamma_s$ and $\gamma_l$ that
are defined by the relations
\begin{equation}
  \label{eq:9}
  \gamma_s = \frac{P_s}{P_m - P_b} \, , \quad \quad  \gamma_l =
 \frac{ P_l }{P_m - P_b} \, .
\end{equation}
This definition has the advantage that the duration $T$ over which a
runner can sustain a given power $P$ can now be expressed as
\begin{equation}
  \label{eq:11}
  T(P)=
   \left\{ 
\begin{array}{ll}
t_c \exp\left[ -\frac{1}{\gamma_l} \frac{P-P_{m}}{  P_{m}- P_b} \right] & P \le P_{m} \\[1em]
t_c \exp\left[ -\frac{1}{\gamma_s} \frac{P -P_{m}}{
  P_{m}- P_b} \right] & P\ge P_{m}
\end{array}\right.
\end{equation}
or in terms of the relative power $p$ [see Eq.~\eqref{eq:6}] as
\begin{equation}
  \label{eq:15}
  T(p)=  \left\{ 
\begin{array}{ll}
t_c \exp\left[ -\frac{p-1}{\gamma_l} \right] & p\le 1 \\[1em]
t_c \exp\left[ -\frac{p-1}{\gamma_s} \right] & p \ge 1
\end{array}\right. \, .
\end{equation}

The time $T$ over which an average velocity $v$ can be sustained follows now
directly by substituting the nominal running economy function of
Eq.~\eqref{eq:economy} into the above equation, leading to
\begin{equation}
  \label{eq:8}
  T(v) = \left\{ 
\begin{array}{ll}
t_c \exp \left[ \frac{v_m-v}{\gamma_l v_m}  \right] & T\ge t_c \, \text{ or } \, v \le v_m \\[1em]
t_c \exp \left[ \frac{v_m-v}{\gamma_s v_m}  \right] & T\le t_c \, \text{ or } \, v \ge v_m
\end{array}\right. \, .
\end{equation}

The fastest performance time $T(d)$ for a distance $d$ can be obtained from
Eq.~\eqref{eq:8} by setting $v=d/T$ and solving for $T$. The solution
can be expressed as the real branch $W_{-1}(z)$ of the Lambert
W-function which is defined as the (multivalued) inverse of the
function $w\to w e^w$ \cite{Corless:1996fh},
\begin{equation}
  \label{eq:T_of_d}
  T(d) = \left\{ 
    \begin{array}{ll}
      - \frac{d}{\gamma_l v_m} \frac{1}{W_{-1}\left[-\frac{d}{d_c \gamma_l}
      e^{-1/\gamma_l}\right]} \, & \text{ for } \, d \ge d_c\\[1em]
       - \frac{d}{\gamma_s v_m} \frac{1}{W_{-1}\left[-\frac{d}{d_c \gamma_s}
      e^{-1/\gamma_s}\right]} \, & \text{ for } \, d \le d_c
      \end{array}\right. \, ,
\end{equation}
where we have defined the  distance $d_c=v_m t_c$ \footnote{The
  function $W_{-1}(z)$ is real valued for $-1/e \le z <0$, a condition
  which is fulfilled for all distances $d$ that we consider.}. Note that $T(d)$
is continuous at $d=d_c$ with $T(d_c)=t_c$ since $W_{-1}(w
e^w)=w$.

This function $T(d)$ can be used the estimate the model parameters
$v_m$, $t_c$, $\gamma_l$ and $\gamma_s$ by minimizing the relative
quadratic error between $T(d_j)$ and the actual race time over
distance $d_j$ for all races $j=1,\ldots,N$. We shall demonstrate this
explicitly below.  From the race time $T(d)$ we can obtain the mean
race velocity for a distance $d$, given by $\bar v(d) = d/T(d)$. When
we express $\bar v(d)$ relative to $v_m$, we obtain the expression
\begin{equation}
  \label{eq:17}
  \frac{\bar v(d)}{v_m} = \left\{ 
    \begin{array}{ll}
      -\gamma_l W_{-1}\left[ -\frac{1}{\gamma_l} \frac{d}{d_c}
      e^{-1/\gamma_l}\right]  & \text{ for } \, d \ge d_c\\[1em]
      -\gamma_s W_{-1}\left[ -\frac{1}{\gamma_s}
      \frac{d}{d_c} e^{-1/\gamma_s}\right]  & \text{ for } \, d \le d_c
    \end{array}\right. \, ,
\end{equation}
which depends only on the parameter $\gamma_l$ (or $\gamma_s$) in the
long (or short) regime when the distance is measured in units of
$d_c$. This function is shown below in Figs.~\ref{fig:v_mean},
\ref{fig:v_mean_ind} for world records and individual runners, and a
typical range of values for $\gamma_l$ and $\gamma_s$.

In order to compare our model predictions to the often assumed power
law or ``broken power law'' description of running records
\cite{Kennely:1906aa,Riegel:1981ci}, it is useful to perform an
asymptotic expansion of the Lambert function $W_{-1}(z)$ for small
negative $z$. This is justified since for all here considered
distances $d$ and model parameters, the argument of $W_{-1}$ in
Eq.~\eqref{eq:T_of_d} never is smaller than $-0.1$. In this range a very
good approximation (better than $0.4\%$) is given by
$W_{-1}(z)=L_1(z)-L_2(z)+L_2(z)/L_1(z) +
[L_2^2(z)-2L_2(z)]/[2L_1^2(z)] + \ldots$ with $L_1(z)=\log(-z)$ and
$L_2=\log(-\log(-z))$. Defining the re-scaled logarithmic time,
distance and mean velocity variables $\tau=\log(T/t_c)$,
$\delta=\log( d/d_c )$ and $\upsilon=\log(\bar v/v_m)$, for $d\ge d_c$
the time-distance and velocity-distance relations are very well
approximated by
\begin{equation}
  \label{eq:time-distance_rel}
  \tau(\delta) = \delta - \upsilon(\delta) \, , \quad
  \upsilon(\delta) = L\left(\frac{1}{\gamma_l},\delta -\log\gamma_l\right)
\end{equation}
with
\begin{equation*}
  \label{eq:L-function}
   L(x,y) = -\log (x) +\log\left[  x - y +
     \log(x-y) + \frac{\log(x-y)}{x-y} -
     \frac{\log(x-y)\left(\log(x-y)-2\right)}{2(x-y)^2}
   \right] \, .
\end{equation*}
The same relations hold for $d\le d_c$ when $\gamma_l$ is replaced by
$\gamma_s$ in Eq.~\eqref{eq:time-distance_rel}.  Note that the
relation between mean race velocity $\bar v$ and race distance $d$ is
not a power law as assumed in some studies
\cite{Riegel:1981ci,S.:2000fi}. For example, Riegel's formula
corresponds in above notation to $\tau(\delta) = \alpha \delta - L$,
$\upsilon(\delta) = - (\alpha-1) \delta + L$ with a {\it constant} $L$
and an exponent $\alpha$ close to $1.06$.  Our model predicts that
$\alpha=1$ exactly and that the very small deviation from $\alpha=1$,
observed by Riegel and others, is due to a hierarchy of logarithmic
corrections, giving rise to a non-constant $L$. It is interesting to
observe from Eq.~\eqref{eq:time-distance_rel} that the endurance
measuring parameter $\gamma_l$ or $\gamma_s$ is the only quantity
which determines the time to distance and velocity to distance
relations when time is measured in units of $t_c$ and velocity in
units of $v_m$. We note that for the comparison of our model to record performances and personal best performances of individual runners, we always use the exact expressions involving the Lambert W-function.

\subsection*{Interpretation of supplemental power $P_{sup}$, and of $\gamma_l$,
  $\gamma_s$}

The supplemental power defined in Eq.~\eqref{eq:2} can be expressed relative
to the aerobic power reserve $P_m-P_b$ as
\begin{equation}
  \label{eq:excess_power}
  \frac{P_{sup}(T)}{P_m-P_b} = \gamma_s \quad \text{ for }  T\le t_c\, , \quad\quad
  \frac{P_{sup}(T)}{P_m-P_b} = (\gamma_s - \gamma_l) \frac{t_c}{T} + \gamma_l
  \quad \text{ for } T > t_c \, ,
\end{equation}
where we used the definitions of Eq.~\eqref{eq:9}.  The averaged
utilized power during a race of duration $T$ and mean velocity $\bar v(T)$,
given by the inverse of Eq.~\eqref{eq:8}, is determined
by the sum of nominal and supplemental power [see Eq.~\eqref{eq:3}],
\begin{equation}
  \label{eq:total_actual_power}
  P_{max} + P_{sup} =
  \left\{ 
    \begin{array}{ll}
     P_b + \frac{\bar v(T)}{v_m} \left[ 1 + \frac{1}{1/\gamma_s - \log (T/t_c)}  \right] & \text{ for } \, T \le t_c\\[1em]
     P_b + \frac{\bar v(T)}{v_m} \left[ 1 +
      \frac{1+(\gamma_s/\gamma_l-1)t_c/T}{1/\gamma_l-\log (T/t_c)} \right] & \text{ for } \, T > t_c
    \end{array}\right. \, .
\end{equation}
The factors in the square brackets measure the amount by which the
total mean running power deviates from the nominal linear relation
$P_b+\bar v(T)/v_m$ with increasing duration $T$. At the crossover
time $t_c$ the factor has its maximum with a value of
$1+\gamma_s$. Below in Fig.~\ref{fig:Excess_factor}, we shall show
graphs of the duration dependence of these supplemental factors for running
world records, and discuss them in relation to experimental
observations.

\begin{figure}[H]
  \centering
 \includegraphics[width=1.0\textwidth]{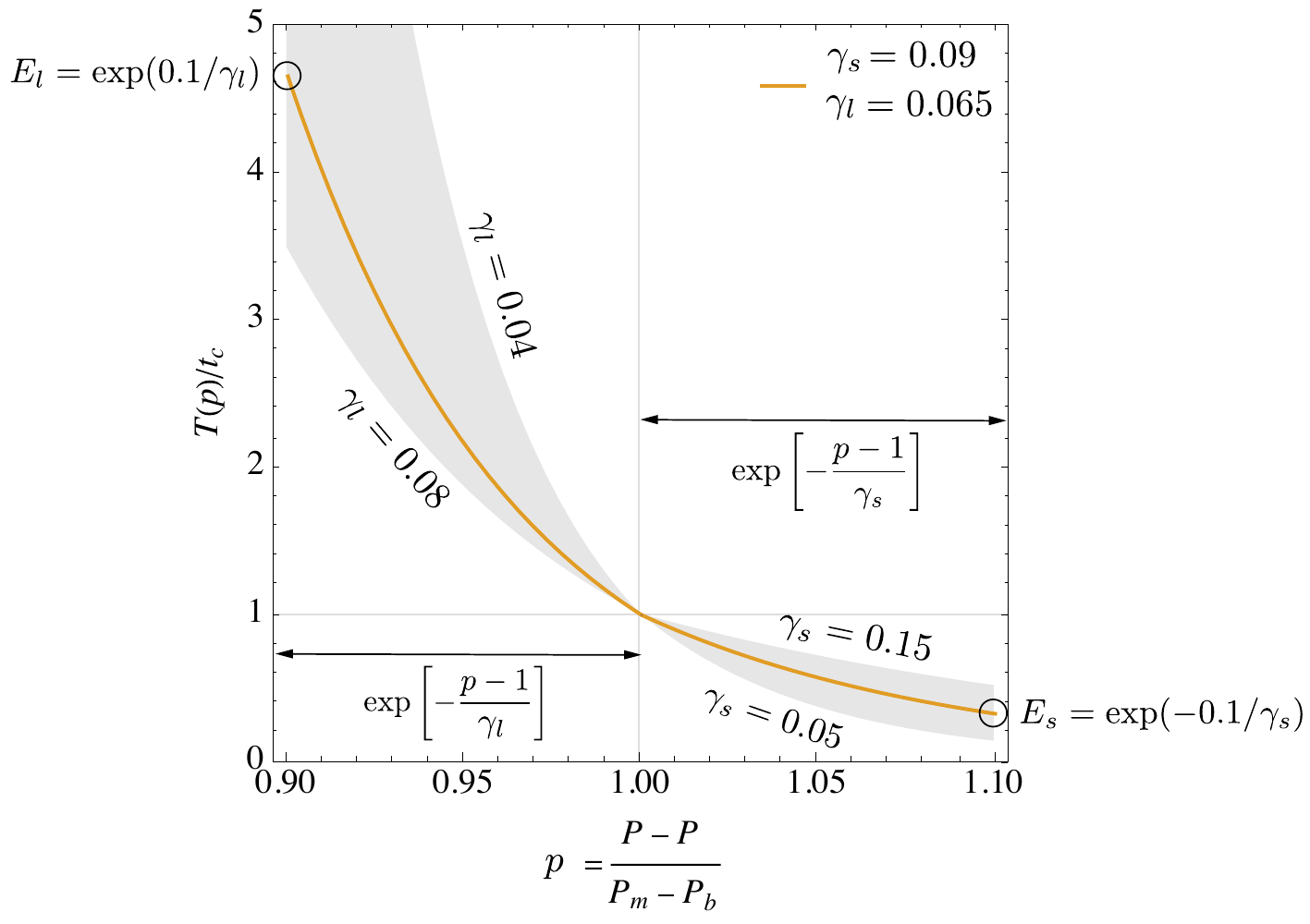}
\caption{{\bf Definition of endurance for long and short duration, $E_{l}$
    and $E_{s}$, respectively, from the duration $T(p)$ over which a
    relative power $p$ can be sustained. Shown is a typical
    range of endurances for long and short duration (gray regions, with
    lower and upper limits for $\gamma_l$ and $\gamma_s$) and an
    example curve that visualizes the definition of $E_{l}$
    and $E_{s}$.}}
  \label{fig:def_endurance}
\end{figure}

\subsection*{Endurance for short and long duration}

The duration $T(p)$ over which a runner can sustain a given relative
power $p$ is shown in Fig.~\ref{fig:def_endurance} for typical
values of the parameters $\gamma_l$ and $\gamma_s$. 
The long and short duration regimes are related by symmetry about the crossover point at $p=1$ due to the same exponential increase (decease) of the duration $T(p)$: Starting from
the crossover power $P_m$, corresponding to $p=1$, the duration
$T(p)$ increases exponentially when the power output is reduced. The rate of this
increase is controlled by the exponent $\gamma_l$. Therefore we define an endurance for {\it long} duration as
$E_l=\exp(0.1 /\gamma_l)$ so that the duration over which a runner can
maintain $90\%$ ($p=0.90$) of crossover power is given by
$T=t_c E_l$. Hence a smaller $\gamma_l$ corresponds to better
endurance.  Similarly, one can ask what parameter range for $\gamma_s$
yields a better performance on shorter distances below the crossover
distance $d_c$. Since in this short duration range one has
$p>1$, the exponential dependence of $T(p)$ yields an increasing
duration with increasing $\gamma_s$. An endurance for {\it short} duration can 
hence be defined as $E_{s}=\exp(-0.1 /\gamma_s)$ so that a runner can
sustain $110\%$ of crossover power for a duration
$T=t_c E_{s}$. Opposite to the long duration regime, here a larger
$\gamma_s$ corresponds to a better endurance. The choice of $90\%$
and $110\%$ of crossover power is arbitrary, and other sub- and
supra-maximal values could be chosen to define endurances without any
qualitative difference in interpretation. We shall come back to these
endurance measures when we discuss personalized characteristic race paces.

\subsection*{Estimation of physiological model parameters}

Our model depends on the four independent parameters $v_m$, $t_c$,
$\gamma_s$ and $\gamma_l$ that characterize a group of runners (for
example world record holders) or individual runners.  Otherwise our
model is universal in the sense that it contains no additional fixed
parameters or constants. The four parameters can be estimated from a
given set of results (distance and time) from exercise performed at
maximal intensity, i.e., races. These sets can be
either records, like world records, involving a group of different runners or
personal records (best performances) from individual runners.  To
check the accuracy of our model and to compute the model
parameters, we minimize numerically the sum of the squared differences
between the actual race time and the one predicted by
Eq.~\eqref{eq:T_of_d} for all results in a given set. This method
will be used to reconstruct individual physiological profiles (running
economy and endurance) from race performances in Application 1 below.

\subsection*{Prediction of race times and characteristic paces for given times and distances}

Once the model parameters for a given set of performance results have been
determined, the model can be applied to compute a number of
interesting quantities that could guide racing and training of a runner.
For example, by comparing the time difference between the actual race
time and the model's prediction for all raced distances, preferred or
optimal distances for a runner can be identified. For distances that
have not been raced before, or only prior to a newly focused training
program, the formula of Eq.~\eqref{eq:T_of_d}, or its approximative version in
Eq.~\eqref{eq:time-distance_rel}, can be used to predict racing times.

Another application of our model is the estimation of characteristic
velocities that correspond to a prescribed relative power output
$\hat p$, measured in percent of aerobic power reserve that is
available over a given duration. Generally, running velocities $v$
in training units depend on the purpose of the training session and
hence on duration $T$ or distance $d$ of the workout intervals.
Suppose that a runner trains at a relative power $\hat p$. This
relative power relates the target power output $P(v)$ to the maximal
power above basal power, $P_{max}(T)-P_b$, that can be maintained for
the duration $T$ by the relation
\begin{equation}
  \label{eq:7}
  P(v) = \hat p ( P_{max}(T)-P_b ) +P_b \, .
\end{equation}
Note that we define here the target power output not relative to the
absolute crossover power but relative to the maximal aerobic
power that can be sustained over time $T$. This is a natural choice
since for a workout of duration $T$, the maximum power that can be
maintained over that time is only $P_{max}(T)$.  Let us assume that a runner
would like to perform a continuous run over a time $T$ at an
intensity $\hat p$, e.g., at $90\%$ ($\hat p=0.9$) of 
maximally possible intensity over that time $T$. Then Eq.~\eqref{eq:7} 
determines under these conditions the velocity
$v$ for the run. An important observation is that the
solution of Eq.~\eqref{eq:7} is independent of both $P_b$ and $P_m$.
In fact, it can be expressed as
\begin{equation}
  \label{eq:v_trai_T}
  v(\hat p,T) = \left\{ 
\begin{array}{ll}
\hat p \, v_m \left[ 1-   \gamma_l  \log(T/t_c) \right] & T \ge t_c \\[.5em]
\hat p \, v_m \left[ 1 - \gamma_s \log(T/t_c) \right] & T \le t_c
\end{array}\right. \, .
\end{equation}
Note that for an intensity of $\hat p=1$ over a time $T=t_c$ one has
$v=v_m$, i.e., the velocity $v_m$ corresponds to the crossover
power, as expected. When instead of time the distance of the
run is fixed, a similar expression for the 
velocity can be derived. Setting $T=d/v$ in Eq.~\eqref{eq:7} and
solving for $v$, one finds
\begin{equation}
  \label{eq:v_trai_d}
  v(\hat p,d) = \left\{ 
    \begin{array}{ll}
      - \hat p \, v_m \gamma_l W_{-1}\left[ - \frac{1}{\hat p \, \gamma_l} \frac{d}{d_c}
      e^{-1/\gamma_l}\right]  & \text{ for } \, d \ge d_c\\[1em]
      - \hat p \, v_m \gamma_s W_{-1}\left[ -\frac{1}{\hat p \, \gamma_s}
      \frac{d}{d_c} e^{-1/\gamma_s}\right]  & \text{ for } \, d \le d_c

    \end{array}\right. \, ,
\end{equation}
where again $d_c=v_m t_c$.  For the intensity $\hat p =1$ this result
corresponds to the race velocity of Eq.~\eqref{eq:17}. It is
important to stress that running economy and endurance both depend on
the absolute values for basal and crossover power, $P_b$ and $P_m$, but race times and paces are determined only by the physiological parameters $v_m$, $\gamma_l$,
$\gamma_s$ and $t_c$.  In Application 2 below we demonstrate the
dependence of race paces on the  physiological parameters
of an athlete.

\section*{Results}

\subsection*{Physiological model parameters from records}

Previously, accurate models for running performance have been based on
a combination of empirical data descriptions and
underlying physiological processes, or they employed at least some
empirical correction factors. Data like world record
performances contain very useful information about maximized
physiological response, and can be used to validate
theoretical models that have been derived entirely from bio-energetic considerations.  Our model fulfills this
requirement, and in this section we shall validate its accuracy by comparing
it to various record performances.

World and other records have been analyzed before and found to follow
an approximate power law. However, the exponent of this power law
shows variations with gender and distance which renders its
universality and general applicability questionable.  Also, there is
no physiological foundation for a simple power law. In fact, the
existence of a crossover velocity $v_m$ implies different scaling of
performances below and above this velocity due to distinct
physiological and bio-energetic processes involved.
 
We have analyzed record performances for eight distances, from 1000m
to the marathon, for world records (current, 2000, 1990, and 1980),
current European records, and current national records (USA, Germany)
see Tab. \ref{tab_records_male} for male records, and Tab.
\ref{tab_records_female} for female records. Following the method
described in the previous  section, we have estimated the
parameters of our model for each group of records.  The resulting parameters
$t_c$, $v_m$, $\gamma_s$, and $\gamma_l$ together with the endurances
$E_s$ and $E_l$ are summarized in Tabs. \ref{tab_records_male},
\ref{tab_records_female}. 
The mean relative error between our model prediction and the VDOT prediction for the race times for 13 distances between 1000m and the marathon are  $0.15\%$, $0.11\%$, and $0.18\%$ for  VDOT=40, 60, and 80, respectively.
These small errors suggest that the race times predicted by the VDOT model are mutually consistent. This presumably reflects that the times were obtain from a mathematical model that is based on physiological observations made by Daniels among well trained and elite runners.

The tables \ref{tab_vdot40_paces},
\ref{tab_vdot60_paces} and \ref{tab_vdot80_paces}
show also the actual race times
$T$ and the times $T_\text{model}$ predicted by our model
Eq.~\eqref{eq:T_of_d}, and the corresponding relative errors in
percent. 

A number of interesting observations can be made from the results: There is a high level of
agreement between actual and predicted times with the relative error
being larger than $1\%$ only for a single event (Half-marathon, WR
1980) for male records, and four events for female records. The mean
of the absolute value of the relative error is always smaller than
$1\%$ with the exception of the female WR from 1990 where it is
$1.05\%$. For the male WR a decrease of the absolute value of the
relative error from 1980 to today can be observed, indicating an
increasing optimization towards the maximally possible performance
(within current level of technology and training methods) that is
described by our model. Hence, the record times have become more
consistent with our model over time which might be also due to an
increasing number of attempts to achieve best possible
performances. A similar observation is made for the female WR
from 1990 to 2000. However, from the 2000 WR some results (Chinese
runner's results for 1500m, 3.000m, 5.000m, and 10.000m) have been
excluded due to the use of performance-enhancing drugs
\cite{Guardian}, and the current WR for 1500m and 10.000m are also
controversial \cite{theguardian_2017_08_04}. For the latter two distances
our model predicts more than $0.5\%$ slower times than actually
raced. It is interesting to observe that our predictions are very
sensitive to exceptional performances for a particular distance
compared to the other distances, and hence is able to identify
suspicious race results. Due to the women's shorter history of endurance
running, the female world records for 1980 are less consistent than
more recent records and hence  have been excluded them from our analysis.

It also instructive to compare the physiological model parameters
obtained from the record performances. For the male records, the obtained values
for $t_c$ vary between five and six minutes, which is in very good agreement with
laboratory testing \cite{billat1998high}. However, for female records,
we observe a larger variation in $t_c$ with values around 10min being
not unusual. However, in cases with such long $t_c$ the crossover
velocity $v_m$ is reduced proportionally. The 
endurance parameter $E_l$ for long distances varies between 5 and 6 for male records,
implying that $90\%$ of maximal aerobic power can be maintained for a
duration between approximately 25min and 36min, for the values of
$t_c$ observed here. For female records, the endurance parameter $E_l$ is
significantly larger with variations in an interval of approximately 6 to 8.5,
implying that $90\%$ of maximal aerobic power can be maintained for
durations up to 85min.

\begin{figure}[H]
  \centering
 \includegraphics[width=.9\textwidth]{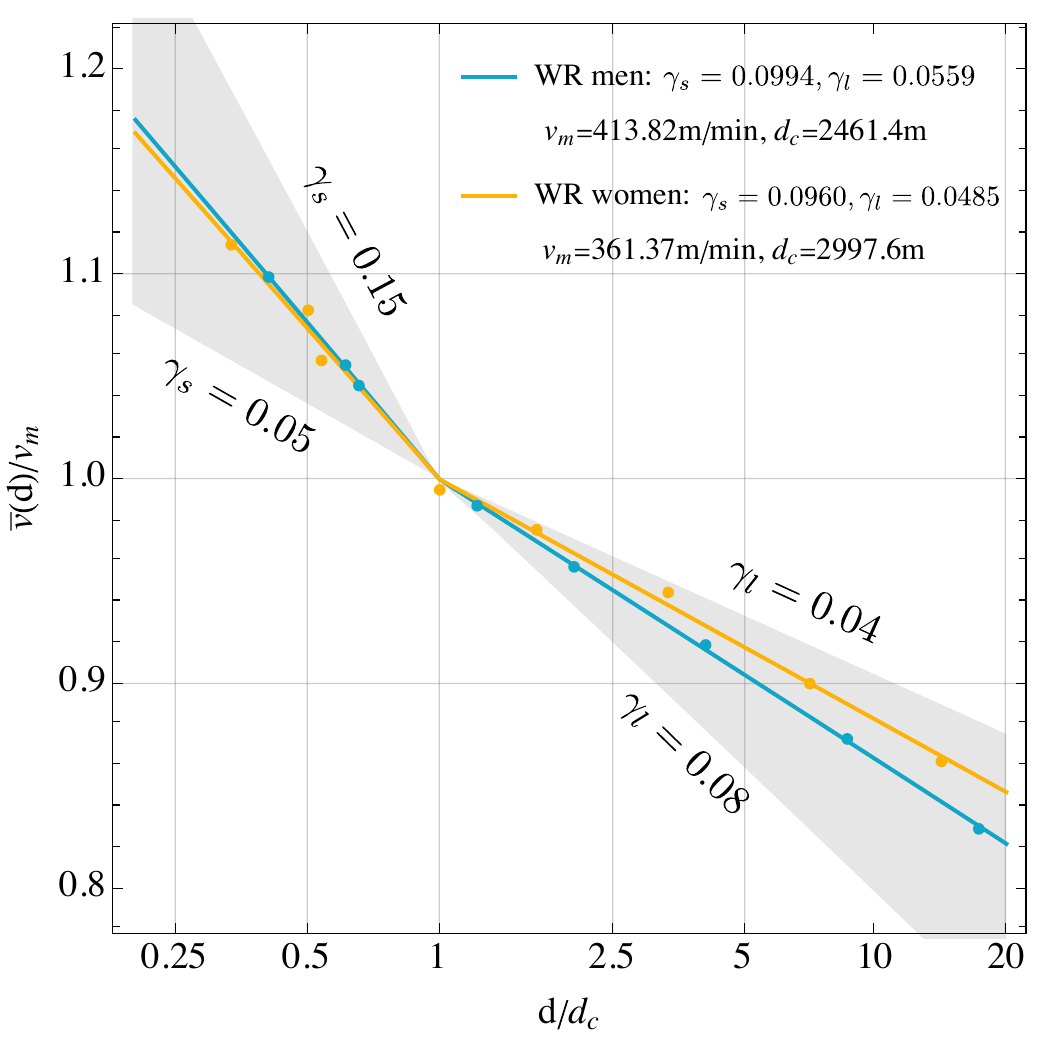}
 \caption{{\bf Mean race velocity $\bar v(d)$ as function
     of race distance. Velocity is re-scaled by $v_m$, and distance $d$
     is re-scaled by $d_c=v_m t_c$. Shown are the male and female world
     records (WR, dots), model prediction from Eq.~\eqref{eq:17}
     (solid lines), and a typically expected maximal range of velocities  (gray regions). Indicated are the lower and upper
     limits of $\gamma_s$ and $\gamma_l$ for these regions. Due to the re-scaling of
     $\bar v(d)$ and $d$, this graph highlights endurance for short and long
     duration, independently of the velocity $v_m$ at maximal
     aerobic power.}}
  \label{fig:v_mean}
\end{figure}

The impact of endurance alone on running performances can be
highlighted by measuring the mean race velocity $\bar v(d)$ in units
of the crossover velocity $v_m$ and the race distance $d$ in units of
the crossover distance $d_c=v_m t_c$. The resulting relation between
$\bar v(d)/v_m$ and $d/d_c$ is shown in Fig.~\ref{fig:v_mean} for the
current world records. Our model predicts that this relation depends
only on the endurance parameters $\gamma_l$ and $\gamma_s$, see
Eq.~\eqref{eq:17}. The corresponding model curves are also plotted in
Fig.~\ref{fig:v_mean}, showing good agreement with the data from world
records. The better endurance of women for distances longer than the
crossover distance $d_c$ is clearly visible.  The gray cones in the
figure indicate the range of endurance parameters that could
potentially be realized in practice by runners from the recreational
to the elite level with suitable event specialization. This type of
visualization of race performances allows one to evaluate runner's
endurance independently of their maximal aerobic power and running
economy which are described by the parameters $v_m$ and $t_c$.

\begin{table}[H]
\begin{adjustwidth}{-1in}{0in}
\centering
\caption{
{\bf Race times and model parameters for various male running records}.}
\addtolength{\tabcolsep}{-4pt}
\begin{tabular}{|L|rrr|rrr|rrr|}
  \hline
 \text{Record} &   \multicolumn{3}{c|}{\text{WR men}}    &    \multicolumn{3}{c|}{\text{WR 2000 men}}  &
   \multicolumn{3}{c|}{\text{WR 1990 men}}    \\
  \hline
 t_c \text{[min]} &   & \text{   5.95} &   &   & \text{   5.50} &   &   & \text{   5.90} &   \\
 v_m \text{[m/min]} &   & \text{ 413.82} &   &   & \text{ 417.07} &   &   & \text{ 405.00} &   \\
 \text{100 }\gamma_s &   & \text{   9.94} &   &   & \text{   9.87} &   &   & \text{  11.76} &   \\
 \text{100 }\gamma_l &   & \text{   5.59} &   &   & \text{   6.19} &   &   & \text{   5.93} &   \\
 E_{\text{s}} &   & \text{   0.37} &   &   & \text{   0.36} &   &   & \text{   0.43} &   \\
 E_l &   & \text{   5.98} &   &   & \text{   5.04} &   &   & \text{
                                                             5.41} &
  \\ \hline
  \text{distance} &  $T$ & $T_\text{model}$ &  $\%$ &  $T$ & $T_\text{model}$ &  $\%$ &  $T$ & $T_\text{model}$ &  $\%$  \\
\hline
 1000 & \text{  02:11.96} & \text{  02:11.94} & -0.02 & \text{  02:11.96} & \text{  02:11.94} & -0.02 & \text{  02:12.80} & \text{ 
   02:12.82} & +0.01 \\
 1500 & \text{  03:26.00} & \text{  03:26.24} & +0.12 & \text{  03:26.00} & \text{  03:26.24} & +0.12 & \text{  03:29.46} & \text{ 
   03:29.26} & -0.10 \\
 1609.34 & \text{  03:43.13} & \text{  03:42.91} & -0.10 & \text{  03:43.13} & \text{  03:42.91} & -0.10 & \text{  03:46.32} & \text{ 
   03:46.50} & +0.08 \\
 3000 & \text{  07:20.67} & \text{  07:20.13} & -0.12 & \text{  07:20.67} & \text{  07:19.38} & -0.29 & \text{  07:29.45} & \text{ 
   07:30.88} & +0.32 \\
 5000 & \text{  12:37.35} & \text{  12:36.76} & -0.08 & \text{  12:39.36} & \text{  12:38.37} & -0.13 & \text{  12:58.39} & \text{ 
   12:56.88} & -0.19 \\
 10000 & \text{  26:17.53} & \text{  26:21.56} & +0.26 & \text{  26:22.75} & \text{  26:33.98} & +0.71 & \text{  27:08.23} & \text{ 
   27:08.68} & +0.03 \\
 21097.5 & \text{  58:23.00} & \text{  58:27.03} & +0.12 & \text{  59:22.00} & \text{  59:18.82} & -0.09 & \text{1:00:46.00} &
   \text{1:00:25.03} & -0.58 \\
 42195 & \text{2:02:57.00} & \text{2:02:44.26} & -0.17 & \text{2:05:42.00} & \text{2:05:26.57} & -0.20 & \text{2:06:50.00} &
   \text{2:07:21.71} & +0.42 \\ \hline
 \text{mean} & \text{} & \text{} & \text{ 0.12} & \text{} & \text{} &
                                                                      \text{ 0.21} & \text{} & \text{} & \text{ 0.22} \\
  \hline
  \hline
 \text{Record} &   \multicolumn{3}{c|}{\text{WR 1980 men}}    &
                                                                 \multicolumn{3}{c|}{\text{US
                                                                 men}}  &
   \multicolumn{3}{c|}{\text{EU men}}    \\
  \hline
  t_c \text{[min]} &   & \text{   5.26} &   &   & \text{   6.08} &   &   & \text{   4.97} &   \\
 v_m \text{[m/min]} &   & \text{ 405.27} &   &   & \text{ 406.06} &   &   & \text{ 412.81} &   \\
 \text{100 }\gamma_s &   & \text{  12.74} &   &   & \text{  10.35} &   &   & \text{  12.24} &   \\
 \text{100 }\gamma_l &   & \text{   6.21} &   &   & \text{   5.67} &   &   & \text{   5.76} &   \\
 E_{\text{s}} &   & \text{   0.46} &   &   & \text{   0.38} &   &   & \text{   0.44} &   \\
 E_l &   & \text{   5.00} &   &   & \text{   5.83} &   &   & \text{
                                                             5.67} &
  \\
  \hline
  \text{distance} &  $T$ & $T_\text{model}$ &  $\%$ &  $T$ & $T_\text{model}$ &  $\%$ &  $T$ & $T_\text{model}$ &  $\%$  \\
\hline
 1000 & \text{  02:13.40} & \text{  02:13.41} & +0.01 & \text{  02:13.90} & \text{  02:13.87} & -0.02 & \text{  02:12.18} & \text{ 
   02:12.17} & -0.01 \\
 1500 & \text{  03:31.36} & \text{  03:31.26} & -0.05 & \text{  03:29.30} & \text{  03:29.61} & +0.15 & \text{  03:28.81} & \text{ 
   03:28.90} & +0.04 \\
 1609.34 & \text{  03:48.80} & \text{  03:48.89} & +0.04 & \text{  03:46.91} & \text{  03:46.62} & -0.13 & \text{  03:46.32} & \text{ 
   03:46.24} & -0.04 \\
 3000 & \text{  07:32.10} & \text{  07:34.44} & +0.52 & \text{  07:29.00} & \text{  07:28.52} & -0.11 & \text{  07:26.62} & \text{ 
   07:26.39} & -0.05 \\
 5000 & \text{  13:08.40} & \text{  13:04.65} & -0.48 & \text{  12:53.60} & \text{  12:51.55} & -0.26 & \text{  12:49.71} & \text{ 
   12:48.61} & -0.14 \\
 10000 & \text{  27:22.47} & \text{  27:30.09} & +0.46 & \text{  26:44.36} & \text{  26:53.60} & +0.58 & \text{  26:46.57} & \text{ 
   26:49.72} & +0.20 \\
 21097.5 & \text{1:02:16.00} & \text{1:01:26.52} & -1.32 & \text{  59:43.00} & \text{  59:41.08} & -0.05 & \text{  59:32.00} & \text{ 
   59:38.51} & +0.18 \\
 42195 & \text{2:09:01.00} & \text{2:10:02.03} & +0.79 & \text{2:05:38.00} & \text{2:05:26.28} & -0.16 & \text{2:05:48.00} &
   \text{2:05:34.12} & -0.18 \\ \hline
 \text{mean} & \text{} & \text{} & \text{ 0.46} & \text{} & \text{} &
                                                                      \text{ 0.18} & \text{} & \text{} & \text{ 0.11} \\
  \hline
\end{tabular}
\hspace*{\fill}

\begin{tabular}{|L|rrr|}
  \hline
 \text{Record} &   \multicolumn{3}{c|}{\text{GER men}}    \\
  \hline
 t_c \text{[min]}&   & \text{   4.79} &   \\
 v_m \text{[m/min]}&   & \text{ 411.05} &    \\
 \text{100 }\gamma_s &   & \text{  11.22} &    \\
 \text{100 }\gamma_l  &   & \text{   6.11} &    \\
 E_{\text{s}} &   & \text{   0.41} &   \\
 E_l &   & \text{   5.14} &   \\
  \hline
  \text{distance} &  $T$ & $T_\text{model}$ &  $\%$  \\
\hline
 1000 & \text{  02:14.53} & \text{  02:14.52} & -0.01 \\
 1500 & \text{  03:31.58} & \text{  03:31.71} & +0.06 \\
 1609.34 & \text{  03:49.22} & \text{  03:49.10} & -0.05 \\
 3000 & \text{  07:30.50} & \text{  07:30.28} & -0.05 \\
 5000 & \text{  12:54.70} & \text{  12:57.09} & +0.31 \\
 10000 & \text{  27:21.53} & \text{  27:13.07} & -0.52 \\
 21097.5 & \text{1:00:34.00} & \text{1:00:45.41} & +0.31 \\
 42195 & \text{2:08:33.00} & \text{2:08:28.19} & -0.06 \\ \hline
 \text{mean} & \text{} & \text{} & \text{ 0.17} \\
  \hline
\end{tabular}
\null\hfill
\addtolength{\tabcolsep}{4pt}
\label{tab_records_male}
\end{adjustwidth}
\end{table}

\begin{table}[H]
\begin{adjustwidth}{-1in}{0in}
\centering
\caption{
{\bf Race times and model parameters for various female records}. $^\dagger$ For the women WR of 2000 the result of 
  Chinese runners for the distances 1500m, 3000m, 5000m and 10000m have
  been excluded due to use of performance-enhancing drugs \cite{Guardian}.}
\addtolength{\tabcolsep}{-4pt}  
\begin{tabular}{|L|rrr|rrr|rrr|}
  \hline
 \text{Record} &   \multicolumn{3}{c|}{\text{WR women}}  &
   \multicolumn{3}{c|}{\text{WR 2000 women$^\dagger$}} &
                                                         \multicolumn{3}{c|}{\text{WR
                                                         1990 women}}\\
  \hline
 t_c \text{[min]} &     & \text{   8.30} &   &   & \text{   10.01} &  & & \text{   5.50}    & \\
 v_m \text{[m/min]}&    & \text{ 361.37} &   &   & \text{ 352.14} &  &   & \text{ 364.74} & \\
 \text{100 }\gamma_s &    & \text{   9.60} &   &   & \text{  10.27} &  &   & \text{  12.13} &\\
 \text{100 }\gamma_l  &    & \text{   4.85} &   &   & \text{   5.53} &   &   & \text{   5.74} & \\
 E_{\text{s}} &     & \text{   0.35} &   &   & \text{   0.38} &  &   & \text{   0.44} &\\
 E_l &    & \text{   7.88} &   &   & \text{ 6.10} &  &   & \text{   5.70} & \\
  \hline
  \text{distance} &  $T$ & $T_\text{model}$ &  $\%$ &  $T$ & $T_\text{model}$ &  $\%$ &  $T$ & $T_\text{model}$ &  $\%$  \\
\hline
 1000 & \text{  02:28.98} & \text{  02:28.78} & -0.13 & \text{  02:28.98} & \text{  02:29.07} & +0.06  & \text{  02:30.67} & \text{  02:30.17} & -0.33\\
 1500 & \text{  03:50.07} & \text{  03:52.05} & +0.86 & \text{  03:52.47} & \text{ 03:52.94} & +0.20 & \text{  03:52.47} & \text{  03:57.27} & +2.06\\
 1609.34 & \text{  04:12.56} & \text{  04:10.70} & -0.74 & \text{  04:12.56} & \text{  04:11.75} & -0.32 & \text{  04:21.68} & \text{  04:16.95} & -1.81\\
 3000 & \text{  08:20.68} & \text{  08:18.13} & -0.51 & \text{  08:21.64} & \text{ 08:21.94} & +0.06 & \text{  08:22.62} & \text{  08:25.93} & +0.66 \\
 5000 & \text{  14:11.15} & \text{  14:12.40} & +0.15 & \text{  14:31.48} & \text{ 14:29.77} & -0.20 & \text{  14:37.33} & \text{  14:31.08} & -0.71\\
 10000 & \text{  29:17.45} & \text{  29:29.07} & +0.66 & \text{  30:13.74} & \text{ 30:14.88} & +0.06 & \text{  30:13.74} & \text{  30:24.18} & +0.58\\
 21097.5 & \text{1:04:51.00} & \text{1:04:50.60} & -0.01 & \text{1:06:40.00} &\text{1:06:56.85} & +0.42 & \text{1:08:32.00} & \text{1:07:34.87} & -1.39\\
 42195 & \text{2:15:25.00} & \text{2:15:00.95} & -0.30 & \text{2:20:43.00} &\text{2:20:18.48} & -0.29 & \text{2:21:06.00} & \text{2:22:16.17} & +0.83 \\ \hline
 \text{mean} & \text{} & \text{} & \text{ 0.42} & \text{} & \text{} & \text{ 0.20} &   \text{} & \text{} & \text{ 1.05} \\
  \hline
  \hline
\text{Record}   &    \multicolumn{3}{c|}{\text{US women}}  &
   \multicolumn{3}{c|}{\text{EU women}}  &  \multicolumn{3}{c|}{\text{GER women}}   \\
  \hline
  t_c \text{[min]} &   & \text{  10.80} &   &   & \text{  10.19} &  &   & \text{   5.87} & \\
 v_m  \text{[m/min]} &   & \text{ 347.42} &   &   & \text{ 351.63} &  &   & \text{ 356.56} & \\
 \text{100 }\gamma_s   &   & \text{   9.39} &   &   & \text{  10.25} &  &   & \text{  13.91} &  \\
 \text{100 }\gamma_l  &   & \text{   5.17} &   &   & \text{   4.63} &  &   & \text{   5.01} &   \\
 E_{\text{s}}   &   & \text{   0.34} &   &   & \text{   0.38} &   &   & \text{   0.49} &  \\
 E_l   &   & \text{   6.92} &   &   & \text{
                                                             8.66} & &   & \text{   7.35} &
  \\
   \hline
  \text{distance} &  $T$ & $T_\text{model}$ &  $\%$ &  $T$ & $T_\text{model}$ &  $\%$ &  $T$ & $T_\text{model}$ &  $\%$  \\
\hline
 1000 & \text{  02:31.80} & \text{  02:32.01} & +0.14 & \text{  02:28.98} & \text{ 
   02:29.08} & +0.07  & \text{  02:30.67} & \text{  02:30.48} & -0.13\\
 1500 & \text{  03:56.29} & \text{  03:56.68} & +0.17 & \text{  03:52.47} & \text{ 
   03:52.92} & +0.20 & \text{  03:57.71} & \text{  03:59.58} & +0.79\\
 1609.34& \text{  04:16.71} & \text{  04:15.62} & -0.42 & \text{  04:12.56} & \text{ 
   04:11.73} & -0.33 & \text{  04:21.59} & \text{  04:19.83} & -0.67 \\
 3000 & \text{  08:25.83} & \text{  08:26.40} & +0.11 & \text{  08:21.42} & \text{ 
   08:21.75} & +0.07 &\text{  08:29.89} & \text{  08:34.62} & +0.93\\
 5000 & \text{  14:38.92} & \text{  14:37.27} & -0.19 & \text{  14:23.75} & \text{ 
   14:27.22} & +0.40  &\text{  14:42.03} & \text{  14:41.99} & -0.00 \\
 10000 & \text{  30:13.17} & \text{  30:24.74} & +0.64 & \text{  29:56.34} & \text{ 
   29:56.03} & -0.02 &\text{  30:57.00} & \text{  30:34.60} & -1.21\\
 21097.5 & \text{1:07:34.00} & \text{1:07:03.57} & -0.75 & \text{1:06:25.00} &
   \text{1:05:40.14} & -1.13 & \text{1:07:58.00} & \text{1:07:25.42} & -0.80 \\
 42195& \text{2:19:36.00} & \text{2:20:00.22} & +0.29 & \text{2:15:25.00} &
   \text{2:16:23.60} & +0.72 &\text{2:19:19.00} & \text{2:20:46.06} & +1.04 \\ \hline
 \text{mean} &\text{} & \text{} & \text{ 0.34} & \text{} & \text{} & \text{ 0.37}  & \text{} & \text{} & \text{ 0.70}\\
  \hline
\end{tabular}
\hspace*{\fill}
\addtolength{\tabcolsep}{4pt}
\label{tab_records_female}
\end{adjustwidth}
\end{table}

\subsection*{Estimate of supplemental power}

\begin{figure}[H]
  \centering
 \includegraphics[width=0.8\textwidth]{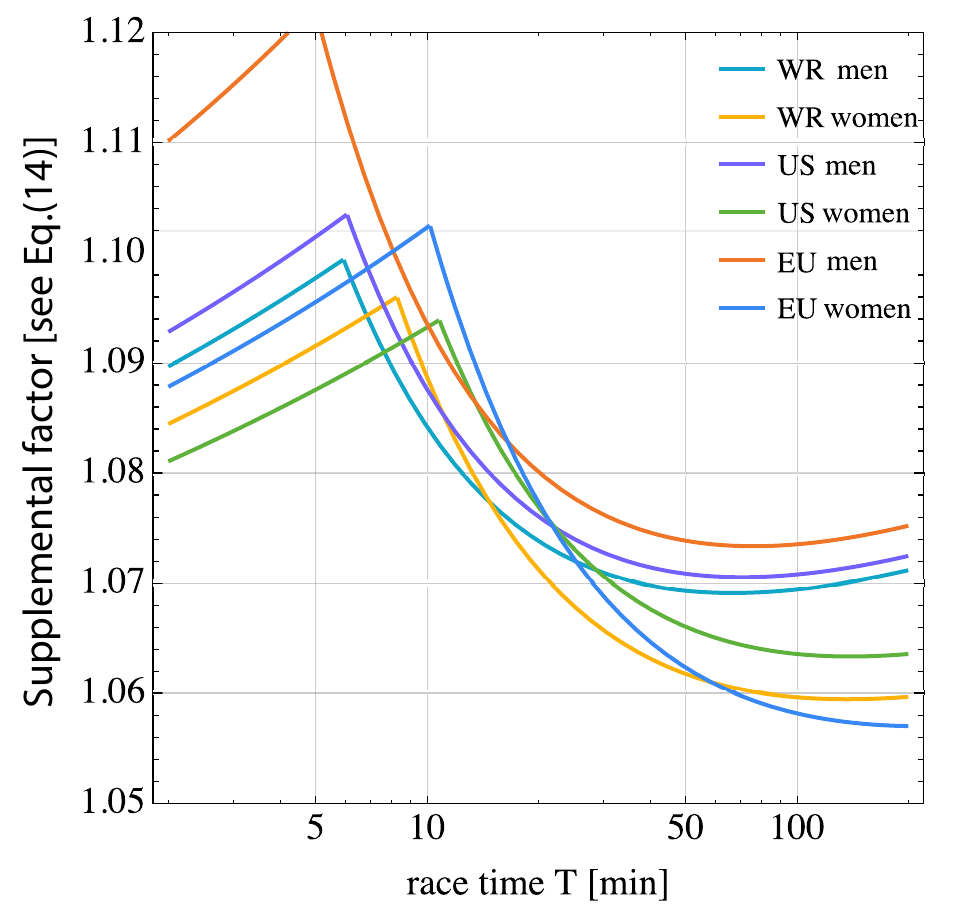}
\caption{{\bf Plot of the supplemental factor of
    Eq.~\eqref{eq:total_actual_power} for as predicted by our model
    for male and female world records (WR), US records (US), and
    European records (EU). The cusp in the curves occurs at the time $t_c$.}}
  \label{fig:Excess_factor}
\end{figure}

We have seen that supplemental power is responsible for a slow logarithmic
decline of racing velocities with distance.  In
Fig.~\ref{fig:Excess_factor} the supplemental factor of
Eq.~\eqref{eq:total_actual_power} (square brackets in this equation)
is plotted for various record performances as function of the race
duration $T$. The variation range of the factor implies a supplemental
power between $\approx$ 6$\%$ and 10$\%$ above the nominal power, with the
European male records (EU men) being an outlier.  The curves have
their maximum at the crossover time $T=t_c$. During supra-maximal
exercise (for times shorter than $t_c$), the oxygen uptake cannot
stabilize and continues to increase until the end of the race
\cite{Gastin:2001aa}. Hence we observe an increasing deviation from
the nominal power with increasing duration. However, at very short
times below about 1 minute, oxygen uptake kinetics limit oxygen supply,
and the energy deficit is compensated by the anaerobic system. After
30 to 60 seconds, the oxygen uptake can reach 90$\%$ of VO$_{2max}$
\cite{Gastin:2001aa}. This short term kinetic effect is not included
in our model.  Above $t_c$, i.e., for sub-maximal velocities, oxygen
uptake stabilizes and the supplemental factor decreases. However, it does
not decrease to one and this is likely related to the fact that the
energy cost of running starts to increase above a nominal linear curve
when the lactate threshold is approached \cite{Batliner:2018aa}. For
even longer race durations, we observe a slight increase in the supplemental
factor that is presumably linked to the increase of the energy cost of
running with increasing distance, as discussed in the
Introduction. For a marathon or a 2 hour run at about 80$\%$
VO$_{2max}$ the supplemental power was measured to be between 5$\%$ and
7$\%$ in terms of oxygen uptake
\cite{Brueckner:1991aa,Brisswalter:2000aa} which is consistent with
our model prediction for $T \sim 120$min. We note that for male
records, the supplemental factor shows a shallow minimum around one hour. For
female records this minimum is displaced to times above two hours.

\subsection*{Application 1: Reconstruction individual physiological profiles}

After we have validated the accuracy of our model against record
performances, we would like to find out if it can be also applied to
individual runners. If that is the case then one could compute from their personal best performances their
individual physiological parameters that characterize their training
state and future performance potential.  The assessment of the
training state of an individual is important not only for performance
optimization but also beyond competitive athletics for the monitoring
of the health status of recreational runners.

There have been performance models developed for
individual runners. A popular model is the so-called VDOT model by Daniels \cite{Daniels3Ed}. This model and
other approaches employ maximal oxygen uptake as single factor
determining performance
\cite{conley1980running,williams1987relationship} and this parameter
is then used to determine the training state and to predict running
performances. A notable exception is the model Peronnet and Thibault
which has been also applied to individual runners
\cite{Peronnet:1991ix}.  It turns out that their model yields
comparable but somewhat larger errors than the present
model. Partially, this might be due their model's assumption that the
energy cost of running and the crossover time $t_c$ would be identical
for all runners. Other physiological factors that determine an
individual's performance include blood lactate concentration, and the
anaerobic threshold. However, these parameters require laboratory
measurements that are not always available, particularly on
sufficiently short time intervals and for recreational athletes.

With the advent of large online databases for personal best
performances, it becomes possible to probe the accuracy of performance
models for a large set of individual athletes. Similar to our analysis of running
records, our model predictions for individual runners can be validated
through comparison with their personal best performances. 
First we reconstruct running economy and endurance profiles of an individual
runner from personal best performances for a few race distances
and then estimate projected race times for other distances and also some
characteristic paces. This eliminates physiological uncertainties that
result from the use of universal, typical physiological parameters in
previous models. In fact, the present model provides a general scheme that can
be applied to any endurance runner over a range of distances and it is
not based on observations made for only a small sample of trained
athletes. Our approach also yields individual relative intensities, in percent of
the aerobic power reserve $P_m-P_b$, at which a runner performs
races. This is important for the relative use of fat and carbohydrate
as fuels, and hence the total carbohydrate consumption for a given
race distance. 

In the following, we apply our model to personal best performances of
British runners that are available online in the database {\tt
  www.thepowerof10.info} \cite{PowerOf10}. As a first test of our
model for individual runners, we have considered the personal bests of
the top nine male and female marathon runners from this database,
according to the 2015 ranking. Their personal best times for seven
distances from 800m to the marathon are summarized in Tables
\ref{tab_marathoners_male_time}, \ref{tab_marathoners_female_time}.
With the same methodology that we used for running records above, we
obtain the four model parameters for each runner that are also listed
in the Tables. From these parameters we compute the predicted race
times.  We find that the agreement between the predicted and actual
race times are the most accurate to date, with an average mean error
of less than 1$\%$ for each individual runner for all seven distances,
see Tables \ref{tab_marathoners_male_time},
\ref{tab_marathoners_female_time}.  This suggests that our model can
describe the running performance of individual runners with reliable
accuracy. The slightly larger mean error for individuals than for
groups of runners (record holders) appears natural since an individual
runner can hardly reach optimized performance for all distances.
When analyzing personal bests of an individual runner one should also realize that the best times on various distances have been probably obtained over a large time span of many years. Especially at the beginning of the career of a runner, when he races predominantly shorter distances, performance might not be optimal. Alternatively, one could consider only best performances obtained within a short time interval like a year which however limits presumably the available distances.

\begin{figure}[H]
  \centering
  \includegraphics[width=.8\textwidth]{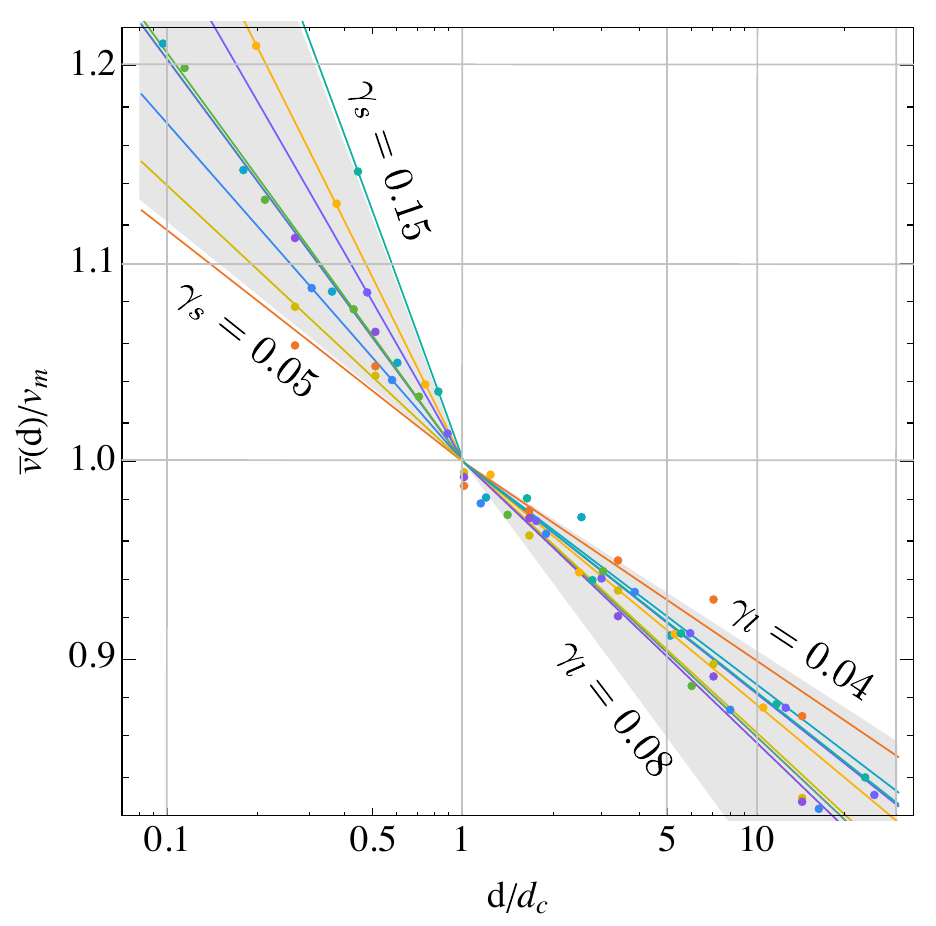}
  \includegraphics[width=.8\textwidth]{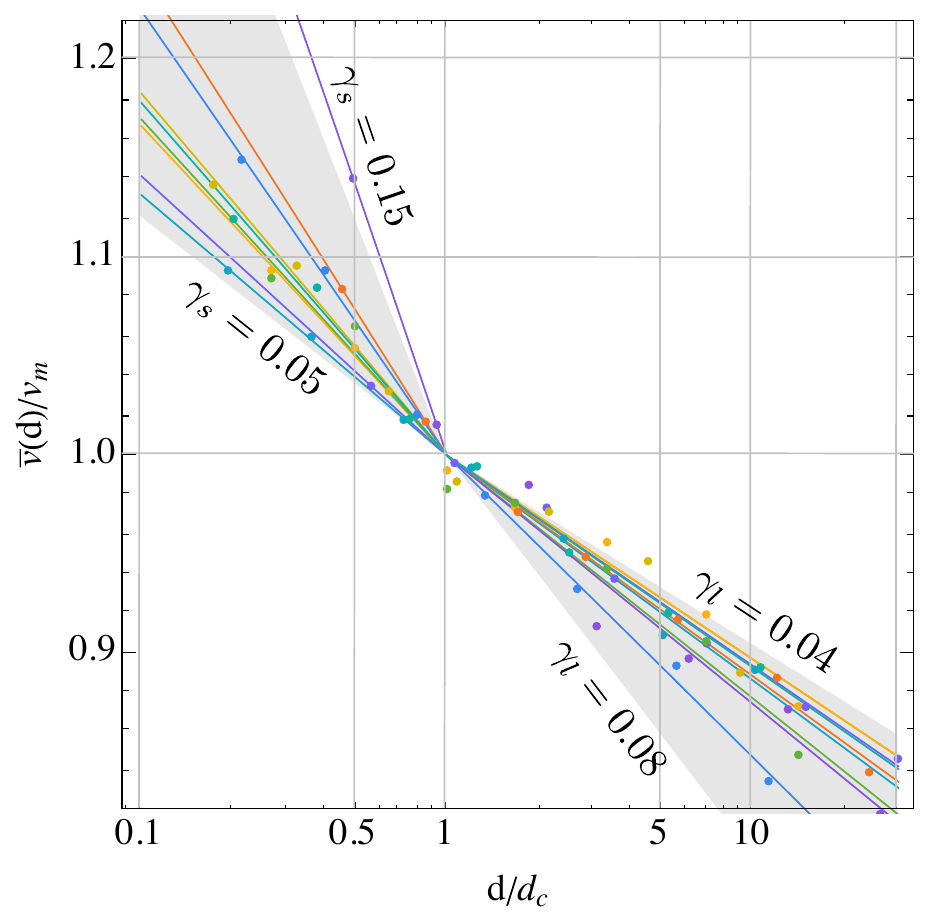}
 \caption{{\bf Same visualization of endurance as in
     Fig.~\ref{fig:v_mean} but for individual male (top) and female
     (bottom) runners, see
     Tabs.~\ref{tab_marathoners_male_time}, \ref{tab_marathoners_female_time}. Colors label different runners.}}
  \label{fig:v_mean_ind}
\end{figure}

Hence the individual variations of the parameters $t_c$ and $v_m$ can be
large but they are strongly correlated. This suggests that $t_c$ gives
a rather precise estimate of the time over which a runner can sustain
the velocity $v_m$ which, however, can deviate slightly from the
actual  velocity at VO$_{2max}$, depending on the available personal
best performances in the vicinity of this crossover point.   In order to measure
individual endurances independently of aerobic capacity, we have
computed and plotted the relation between the re-scaled race velocity
$\bar v(d)/v_m$ and distance $d/d_c$ in analogy to our analysis of
running records, see Fig.~\ref{fig:v_mean_ind}. Two important
observations can be made from this graph: (1) For each individual
runner, there are two distinct relations between velocity and distance
above and below the crossover velocity $v_m$ and distance $d_c$. (2)
Even within the group of top UK marathon runners, there is a large
variation in endurances as quantified by the different slopes of the
re-scaled velocity-distance curves and the parameters $\gamma_s$ and
$\gamma_l$. They gray cones of expected maximal variations shown in
Fig.~\ref{fig:v_mean_ind} are almost completely covered by the
performances of the studied runners.

Our findings show that individual performances do not follow a unique
power law as suggested, for example, by Riegel's formula. There are
more complex variations of physiological metrics among runners and
those have to be taken into account for describing and predicting
accurately performances and presumably optimal training.  Our computational
approach reveals the physiological parameters that determine
individual performance and explains how they can be used in
praxis to guide training and racing.

\begin{table}[H]
\begin{adjustwidth}{-1in}{0in}
\vspace*{-0.6cm}
\centering
\caption{
{\bf Personal best times and model parameters for individuals (Leading male
  marathon runners from UK, ranking 2015, {\tt http://www.thepowerof10.info/rankings}).}}
\addtolength{\tabcolsep}{-4pt}  
\begin{tabular}{|L|rrr|rrr|rrr|}
\hline
\text{Runner} &   & 01 &   &   & 02 &   &   & 03 &   \\ \hline
 t_c \text{[min]}&   & \text{  23.84} &   &   & \text{  11.28} &   &   & \text{   4.57} &   \\
 v_m \text{[m/min]}&   & \text{ 353.28} &   &   & \text{ 360.35} &   &   & \text{ 373.49} &   \\
 \text{100 }\gamma_s&   & \text{   8.16} &   &   & \text{  11.65} &   &   & \text{  10.26} &   \\
 \text{100 }\gamma_l &   & \text{   4.67} &   &   & \text{   5.07} &   &   & \text{   4.85} &   \\
 E_{\text{s}} &   & \text{   0.29} &   &   & \text{   0.42} &   &   & \text{   0.38} &   \\
 E_l &   & \text{   8.52} &   &   & \text{   7.20} &   &   & \text{   7.86} &   \\
\hline
\text{distance} &  $T$ & $T_\text{model}$ &  $\%$ &  $T$ & $T_\text{model}$ &  $\%$ &  $T$ & $T_\text{model}$ &  $\%$  \\
  \hline
800 & \text{  01:52.08} & \text{  01:52.52} & +0.39 & \text{  01:49.98} & \text{  01:49.94} & -0.04 & \text{  01:58.32} & \text{  01:58.32} & +0.00 \\
 1500 & \text{  03:41.88} & \text{  03:41.06} & -0.37 & \text{  03:40.80} & \text{  03:40.95} & +0.07 & \text{  03:57.48} & \text{  03:57.48} & -0.00 \\
 3000 & \text{  07:48.90} & \text{  07:46.84} & -0.44 & \text{  08:00.48} & \text{  08:00.34} & -0.03 & \text{  08:16.62} & \text{  08:16.24} & -0.08 \\
 5000 & \text{  13:28.32} & \text{  13:31.65} & +0.41 & \text{  13:57.66} & \text{  14:01.83} & +0.50 & \text{  14:13.32} & \text{  14:09.91} & -0.40 \\
 10000 & \text{  28:49.02} & \text{  28:32.80} & -0.94 & \text{  29:23.04} & \text{  29:09.19} & -0.79 & \text{  29:18.48} & \text{  29:26.13} & +0.43 \\
 21097.5 & \text{1:01:25.02} & \text{1:02:32.11} & +1.82 & \text{1:04:07.02} & \text{1:04:12.25} & +0.14 & \text{1:04:30.00} & \text{1:04:49.91} & +0.51 \\
 42195 & \text{2:10:55.02} & \text{2:09:41.73} & -0.93 &
                                                         \text{2:13:40.98} & \text{2:13:52.45} & +0.14 & \text{2:15:51.00} & \text{2:15:11.79} & -0.48 \\ \hline
 \text{mean} & \text{} & \text{} & \text{ 0.76} & \text{} & \text{} & \text{ 0.24} & \text{} & \text{} & \text{ 0.27} \\
  \hline
  \hline
  \text{Runner} &   & 04 &   &   & 05 &   &   & 06 &   \\ \hline
 t_c \text{[min]}&   & \text{  19.88} &   &   & \text{   8.57} &   &   & \text{   6.88} &   \\
 v_m \text{[m/min]}&   & \text{ 357.87} &   &   & \text{ 349.94} &   &   & \text{ 382.82} &   \\
 \text{100 }\gamma_s &   & \text{   8.27} &   &   & \text{   4.84} &   &   & \text{   6.93} &   \\
 \text{100 }\gamma_l &   & \text{   5.70} &   &   & \text{   4.19} &   &   & \text{   5.70} &   \\
 E_{\text{s}} &   & \text{   0.30} &   &   & \text{   0.13} &   &   & \text{   0.24} &   \\
 E_l &   & \text{   5.78} &   &   & \text{  10.86} &   &   & \text{
                                                             5.79} &\\
\hline
\text{distance} &  $T$ & $T_\text{model}$ &  $\%$ &  $T$ & $T_\text{model}$ &  $\%$ &  $T$ & $T_\text{model}$ &  $\%$  \\
  \hline
  800 & \text{  01:51.78} & \text{  01:52.20} & +0.38 & \text{  02:09.48} & \text{  02:08.54} & -0.73 & \text{  01:55.20} & \text{  01:55.20} & +0.00 \\
 1500 & \text{  03:41.94} & \text{  03:40.71} & -0.56 & \text{  04:05.22} & \text{  04:08.44} & +1.31 & \text{  03:45.66} & \text{  03:45.66} & -0.00 \\
 3000 & \text{  07:46.74} & \text{  07:46.78} & +0.01 & \text{  08:40.50} & \text{  08:34.37} & -1.18 & \text{  08:00.12} & \text{  07:53.93} & -1.29 \\
 5000 & \text{  13:31.20} & \text{  13:32.52} & +0.16 & \text{  14:38.58} & \text{  14:36.90} & -0.19 & \text{  13:33.00} & \text{  13:35.29} & +0.28 \\
 10000 & \text{  28:42.18} & \text{  28:31.86} & -0.60 & \text{  30:04.02} & \text{  30:10.06} & +0.33 & \text{  27:57.24} & \text{  28:25.13} & +1.66 \\
 21097.5 & \text{1:02:22.98} & \text{1:03:06.46} & +1.16 & \text{1:04:46.98} & \text{1:05:55.65} & +1.77 & \text{1:03:00.00} & \text{1:03:04.39} & +0.12 \\
 42195 & \text{2:12:57.00} & \text{2:12:10.52} & -0.58 &
                                                         \text{2:18:21.00} & \text{2:16:24.13} & -1.41 & \text{2:13:40.02} & \text{2:12:33.95} & -0.82 \\ \hline
 \text{mean} & \text{} & \text{} & \text{ 0.49} & \text{} & \text{} & \text{ 0.99} & \text{} & \text{} & \text{ 0.60} \\
  \hline
  \hline
\text{Runner} &   & 07 &   &   & 08 &   &   & 09 &   \\ \hline
 t_c \text{[min]}&   & \text{   8.44} &   &   & \text{   8.17} &   &   & \text{   5.28} &   \\
 v_m \text{[m/min]}&   & \text{ 355.63} &   &   & \text{ 367.03} &   &   & \text{ 347.39} &   \\
 \text{100 }\gamma_s &   & \text{   5.72} &   &   & \text{   8.15} &   &   & \text{  15.25} &   \\
 \text{100 }\gamma_l &   & \text{   5.62} &   &   & \text{   5.81} &   &   & \text{   4.82} &   \\
 E_{\text{s}} &   & \text{   0.17} &   &   & \text{   0.29} &   &   & \text{   0.52} &   \\
 E_l &   & \text{   5.93} &   &   & \text{   5.59} &   &   & \text{   7.95} &   \\
\hline
\text{distance} &  $T$ & $T_\text{model}$ &  $\%$ &  $T$ & $T_\text{model}$ &  $\%$ &  $T$ & $T_\text{model}$ &  $\%$  \\
  \hline
800 & \text{  02:05.10} & \text{  02:04.97} & -0.11 & \text{  01:57.42} & \text{  01:57.12} & -0.26 & \text{  02:00.42} & \text{  02:00.42} & +0.00 \\
 1500 & \text{  04:02.40} & \text{  04:02.87} & +0.19 & \text{  03:49.98} & \text{  03:51.04} & +0.46 & \text{  04:10.08} & \text{  04:10.08} & -0.00 \\
 3000 & \text{  08:28.62} & \text{  08:26.15} & -0.49 & \text{  08:14.04} & \text{  08:10.42} & -0.73 & \text{  08:47.70} & \text{  08:51.43} & +0.71 \\
 5000 & \text{  14:35.94} & \text{  14:30.06} & -0.67 & \text{  14:01.02} & \text{  14:04.01} & +0.36 & \text{  15:18.30} & \text{  15:09.93} & -0.91 \\
 10000 & \text{  30:04.02} & \text{  30:17.72} & +0.76 & \text{  29:32.70} & \text{  29:26.30} & -0.36 & \text{  31:30.90} & \text{  31:30.08} & -0.04 \\
 21097.5 & \text{1:06:04.02} & \text{1:07:09.06} & +1.64 & \text{1:04:28.02} & \text{1:05:23.05} & +1.42 & \text{1:09:12.00} & \text{1:09:20.92} & +0.21 \\
 42195 & \text{2:22:55.98} & \text{2:20:56.89} & -1.39 &
                                                         \text{2:18:49.02} & \text{2:17:31.77} & -0.93 & \text{2:24:31.02} & \text{2:24:32.68} & +0.02 \\ \hline
 \text{mean} & \text{} & \text{} & \text{ 0.75} & \text{} & \text{} & \text{ 0.65} & \text{} & \text{} & \text{ 0.27} \\
  \hline
\end{tabular}
\addtolength{\tabcolsep}{4pt}  
\label{tab_marathoners_male_time}
\end{adjustwidth}
\end{table}

\begin{table}
\begin{adjustwidth}{-1in}{0in}
\vspace*{-0.6cm}
\centering
\caption{
{\bf Personal best times and model parameters for individuals (Leading female
  marathon runners from UK, ranking 2015, {\tt http://www.thepowerof10.info/rankings}).}}
\addtolength{\tabcolsep}{-4pt}  
\begin{tabular}{|L|rrr|rrr|rrr|}
  \hline
\text{Runner} &   & 01 &   &   & 02 &   &   & 03 &   \\ \hline
 t_c \text{[min]}&   & \text{  13.01} &   &   & \text{   9.48} &   &   & \text{   3.85} &   \\
 v_m \text{[m/min]}&   & \text{ 319.66} &   &   & \text{ 316.60} &   &   & \text{ 368.12} &   \\
 \text{100 }\gamma_s&   & \text{   5.45} &   &   & \text{   6.80} &   &   & \text{   5.82} &   \\
 \text{100 }\gamma_l &   & \text{   4.70} &   &   & \text{   4.28} &   &   & \text{   4.41} &   \\
 E_{\text{s}} &   & \text{   0.16} &   &   & \text{   0.23} &   &   & \text{   0.18} &   \\
 E_l &   & \text{   8.39} &   &   & \text{  10.34} &   &   & \text{
                                                             9.64} & \\
\hline
\text{distance} &  $T$ & $T_\text{model}$ &  $\%$ &  $T$ & $T_\text{model}$ &  $\%$ &  $T$ & $T_\text{model}$ &  $\%$  \\
  \hline
800 & \text{  02:17.28} & \text{  02:17.17} & -0.08 & \text{  02:18.60} & \text{  02:18.31} & -0.21 & \text{  02:05.94} & \text{  02:05.94} & -0.00 \\
 1500 & \text{  04:25.56} & \text{  04:25.95} & +0.15 & \text{  04:29.58} & \text{  04:30.60} & +0.38 & \text{  04:05.40} & \text{  04:05.12} & -0.11 \\
 3000 & \text{  09:13.08} & \text{  09:12.71} & -0.07 & \text{  09:32.82} & \text{  09:28.54} & -0.75 & \text{  08:22.20} & \text{  08:26.50} & +0.86 \\
 5000 & \text{  15:44.22} & \text{  15:47.11} & +0.31 & \text{  16:13.02} & \text{  16:09.74} & -0.34 & \text{  14:29.10} & \text{  14:25.36} & -0.43 \\
 10000 & \text{  32:39.36} & \text{  32:42.02} & +0.14 & \text{  33:01.98} & \text{  33:23.16} & +1.07 & \text{  30:01.08} & \text{  29:51.82} & -0.51 \\
 21097.5 & \text{1:12:36.00} & \text{1:11:45.67} & -1.16 & \text{1:12:28.02} & \text{1:13:01.34} & +0.77 & \text{1:05:40.02} & \text{1:05:30.03} & -0.25 \\
 42195 & \text{2:28:04.02} & \text{2:29:05.72} & +0.69 &
                                                         \text{2:32:40.02} & \text{2:31:12.50} & -0.96 & \text{2:15:25.02} & \text{2:16:00.78} & +0.44 \\ \hline
 \text{mean} & \text{} & \text{} & \text{ 0.37} & \text{} & \text{} & \text{ 0.64} & \text{} & \text{} & \text{ 0.37} \\
  \hline
\hline
\text{Runner} &   & 04 &   &   & 05 &   &   & 06 &   \\ \hline
 t_c \text{[min]}&   & \text{   9.45} &   &   & \text{   5.92} &   &   & \text{  12.36} &   \\
 v_m \text{[m/min]}&   & \text{ 317.48} &   &   & \text{ 297.47} &   &   & \text{ 303.80} &   \\
 \text{100 }\gamma_s &   & \text{   6.93} &   &   & \text{   9.70} &   &   & \text{   9.00} &   \\
 \text{100 }\gamma_l &   & \text{   5.06} &   &   & \text{   4.62} &   &   & \text{   6.19} &   \\
 E_{\text{s}} &   & \text{   0.24} &   &   & \text{   0.36} &   &   & \text{   0.33} &   \\
 E_l &   & \text{   7.23} &   &   & \text{   8.71} &   &   & \text{   5.02} &   \\
  \hline
\text{distance} &  $T$ & $T_\text{model}$ &  $\%$ &  $T$ & $T_\text{model}$ &  $\%$ &  $T$ & $T_\text{model}$ &  $\%$  \\
  \hline
800 & \text{  02:18.72} & \text{  02:17.68} & -0.75 & \text{  02:28.80} & \text{  02:28.80} & -0.00 & \text{  02:17.40} & \text{  02:17.16} & -0.18 \\
 1500 & \text{  04:26.04} & \text{  04:29.59} & +1.33 & \text{  04:57.42} & \text{  04:57.42} & -0.00 & \text{  04:30.84} & \text{  04:31.69} & +0.31 \\
 3000 & \text{  09:36.72} & \text{  09:26.96} & -1.69 & \text{  10:22.86} & \text{  10:21.12} & -0.28 & \text{  09:40.44} & \text{  09:39.64} & -0.14 \\
 5000 & \text{  16:08.10} & \text{  16:11.38} & +0.34 & \text{  17:43.02} & \text{  17:42.23} & -0.07 & \text{  16:47.82} & \text{  16:46.53} & -0.13 \\
 10000 & \text{  33:24.72} & \text{  33:39.59} & +0.74 & \text{  36:40.02} & \text{  36:42.61} & +0.12 & \text{  35:18.00} & \text{  35:11.88} & -0.29 \\
 21097.5 & \text{1:13:21.00} & \text{1:14:10.86} & +1.13 & \text{1:19:55.02} & \text{1:20:39.07} & +0.92 & \text{1:17:43.02} & \text{1:18:25.22} & +0.90 \\
 42195 & \text{2:36:39.00} & \text{2:34:47.30} & -1.19 &
                                                         \text{2:48:55.98} & \text{2:47:45.34} & -0.70 & \text{2:46:19.02} & \text{2:45:29.11} & -0.50 \\ \hline
 \text{mean} & \text{} & \text{} & \text{ 1.03} & \text{} & \text{} & \text{ 0.30} & \text{} & \text{} & \text{ 0.35} \\
  \hline
\hline
  \text{Runner} &   & 07 &   &   & 08 &   &   & 09 &   \\ \hline
 t_c \text{[min]} &   & \text{  16.50} &   &   & \text{   5.40} &   &   & \text{  14.64} &   \\
 v_m \text{[m/min]}&   & \text{ 281.73} &   &   & \text{ 300.16} &   &   & \text{ 272.55} &   \\
 \text{100 }\gamma_s&   & \text{   7.43} &   &   & \text{  16.76} &   &   & \text{   7.25} &   \\
 \text{100 }\gamma_l &   & \text{   4.28} &   &   & \text{   5.16} &   &   & \text{   4.45} &   \\
 E_{\text{s}} &   & \text{   0.26} &   &   & \text{   0.55} &   &   & \text{   0.25} &   \\
 E_l &   & \text{  10.33} &   &   & \text{   6.93} &   &   & \text{   9.47} &   \\
  \hline
\text{distance} &  $T$ & $T_\text{model}$ &  $\%$ &  $T$ & $T_\text{model}$ &  $\%$ &  $T$ & $T_\text{model}$ &  $\%$  \\
  \hline
 800 & \text{  02:29.82} & \text{  02:29.38} & -0.29 & \text{  02:20.22} & \text{  02:20.22} & -0.00 & \text{  02:37.26} & \text{  02:36.54} & -0.46 \\
 1500 & \text{  04:51.42} & \text{  04:52.95} & +0.52 & \text{  04:55.20} & \text{  04:55.20} & -0.00 & \text{  05:04.32} & \text{  05:06.82} & +0.82 \\
 3000 & \text{  10:18.72} & \text{  10:17.25} & -0.24 & \text{  10:08.70} & \text{  10:20.47} & +1.93 & \text{  10:48.48} & \text{  10:46.06} & -0.37 \\
 5000 & \text{  17:58.98} & \text{  17:48.35} & -0.98 & \text{  18:13.98} & \text{  17:44.85} & -2.66 & \text{  18:26.70} & \text{  18:32.44} & +0.52 \\
 10000 & \text{  36:31.98} & \text{  36:45.40} & +0.61 & \text{  37:07.98} & \text{  36:59.37} & -0.39 & \text{  38:34.98} & \text{  38:19.98} & -0.65 \\
 21097.5 & \text{1:19:07.02} & \text{1:20:19.99} & +1.54 & \text{1:20:39.00} & \text{1:21:45.39} & +1.37 & \text{1:24:06.00} & \text{1:23:55.81} & -0.20 \\
 42195 & \text{2:48:16.02} & \text{2:46:13.19} & -1.22 &
                                                         \text{2:51:46.02} & \text{2:51:06.21} & -0.39 & \text{2:53:25.02} & \text{2:53:58.66} & +0.32 \\ \hline
 \text{mean} & \text{} & \text{} & \text{ 0.77} & \text{} & \text{} & \text{ 0.96} & \text{} & \text{} & \text{ 0.48} \\
    \hline
\end{tabular}
\addtolength{\tabcolsep}{4pt}  
\label{tab_marathoners_female_time}
\end{adjustwidth}
\end{table}

\subsection*{Application 2: Personalized characteristic paces}

We expect that our four parameter model can measure an individual
runner's performance status for distances from 800m to the marathon more
accurately than previous performance models that often assume for all
runners the same (average) values for certain characteristics like
running economy or endurance.  An example for the latter type of
models is the popular VDOT model of J. Daniels which assumes a fixed
running economy and endurance curves for all runners
\cite{Daniels3Ed,OxygenPower}.  Although the VDOT model represents a
good first approximation of characteristic paces based on a single race
performance, the ability to monitor individual performances with more
than just one parameter allows the runner to ascertain a better
understanding of their training status and potential performance.  It then becomes
beneficial to have a model that makes use of larger available data
sets.  In the same way that one may better understand current fitness
by examining relative oxygen consumption at different paces rather
than absolute oxygen consumption, \cite{bassett2000limiting}
developing an approach that makes use of performance over several
races describes an individual runner better than a single race.

Characteristic paces are often defined by the pace that a runner can race
(at current training status) for a prescribed duration or
distance. When the physiological model parameters of a runner are
known from sufficiently many recent race performances, the running
velocities for a prescribed intensity and duration, or intensity and
distance can be computed from Eq.~\eqref{eq:v_trai_T} and
Eq.~\eqref{eq:v_trai_d}, respectively. In the following we consider
race paces for a given duration or distance, corresponding to
$\hat p =1$ in these equations.  In order to compare our model
predictions to the characteristic paces of the VDOT model, we consider three
hypothetical runners that are assumed to have achieved race
performances as predicted by the VDOT model with model parameter
values VDOT=40, 60, and 80. (VDOT can be regarded as an effective
value for VO$_{2max}$, see \cite{Daniels3Ed} for details.) From these
race performances we obtain the four parameters of our model. These
parameters are given in the captions of Tables \ref{tab_vdot40_paces},
\ref{tab_vdot60_paces} and \ref{tab_vdot80_paces}. These tables
provide race paces (time per km) for various distances and durations
specified in the first column. Some of the paces correspond the
specific paces named in the VDOT model, and they are labeled
correspondingly as R-, I-, T- and M-pace. The paces proposed by the
VDOT model are given in the second column. The remaining columns
provide the predictions of our model. The third column lists the paces
as obtained from the values of the four model parameters that result from the hypothetical race performances of the
runner with the given VDOT score. There is agreement within a few seconds per
kilometer. It should be kept in mind that our model, unlike the VDOT
model, does not implement any fixed parameters or constants a priori.
We observe that the fixed parameters of the VDOT model correspond to
rather superior endurance with $\gamma_l\approx 0.05$ for long
distances and average endurance with $\gamma_s \approx 0.09$ for short
distances. As we have seen above, there is substantial variation in
these parameters among individuals. Hence, characteristic paces should also
determined individually. We have modified the endurance parameters
$\gamma_l$ and $\gamma_s$ independently within their typical minimal
and maximal values while keeping $v_m$ and $t_c$ unchanged. The
resulting paces are shown in the last four columns of the tables.  The
fast paces for short distances (1mile and 5min paces) can change up to
$\pm 10$sec/km compared to the original VDOT model which is
substantial.  For the slower paces (for time $t_c$ and longer) the
variation can be even larger with a maximum change for the marathon
pace (M-pace). For a VDOT=40 runner, the M-pace window between slowest
and fastest pace is about 55sec/km, for a VDOT=60 runner it is about
30sec/km and even for a high level runner with VDOT=80 it is still
about 20sec/km. These variations result from  different
endurances, with the crossover speed $v_m$ unchanged. We have also
studied the effect of a modification of the time $t_c$ from the original
VDOT model value which appears rather long with 12 to 13min. The
results are shown in Tables \ref{tab_vdot40_paces_2},
\ref{tab_vdot60_paces_2} and \ref{tab_vdot80_paces_2}. The first three
columns have the same meaning as in the three tables before. The last
four columns list the paces that correspond to a reduction or an
increase of $t_c$ by $10\%$ or $20\%$, respectively. Here we observe a smaller variation by a few seconds around the original paces,
relatively independent of the duration or distance that defines the
pace.  This shows that racing paces are more dependent   
on endurance than on the time over which runners can sustain their
crossover speed at VO$_{2max}$. The reason for that is the exponential
dependence on $\gamma_s$, $\gamma_l$ of the duration $T(p)$ over which
a relative power $p$ can be maintained, independently of $t_c$ and
$v_m$, see Fig.~\ref{fig:def_endurance}.

It is interesting to relate this observation to physiological
parameters that can be measured in the laboratory and have been linked
to endurance capacity, like blood lactate concentration.  It is known
that the running speed at the lactate threshold can improve
independently of VO$_{2max}$ and so can the runner's endurance. Often
the lactate threshold pace is identified with the running velocity
that a runner can race for about 60min. The corresponding paces are
shown in Tabs.~\ref{tab_vdot40_paces}-~\ref{tab_vdot80_paces_2} as
``T-pace''.  The relative intensity or power output in percent of the
aerobic power reserve [see Eq.~\eqref{eq:6}] at the lactate threshold
is given by $p_{LT} = 100[1-\gamma_l \log (60/t_c)]$. For example, for
a recreational runner (with VDOT=40), described by the parameters of
Tab.~\ref{tab_vdot40_paces}, one has $p_{LT}=91.94\%$ for the original
value $\gamma_l=0.051$, while $p_{LT}=93.68\%$ for $\gamma_l=0.04$,
and $p_{LT}=87.35\%$ for $\gamma_l=0.08$. These values appear rather
large when compared to the lactate threshold estimates from current
world records: $p_{LT}=87.08\%$ for male and $p_{LT}=90.41\%$ for
female records. This implies again that the VDOT model assumes a rather
optimized endurance.

\begin{table}[H]
\begin{adjustwidth}{-1in}{0in}
\centering
\caption{
{\bf Paces per km for a runner with VDOT=40 score for different endurances. The original physiological
  parameters are $t_c=12.35$min, $v_m=214.88$m/min, $\gamma_l=0.051$ and
$\gamma_s=0.096$. In last 4 columns the endurances $E_l$ and $E_s$ are given only when
they are different from the original values.}}
\addtolength{\tabcolsep}{-4pt}  
\begin{tabular}{|c|r|r|r|r|r|r|}
  \hline
  \text{pace at} & Ref.~\cite{Daniels3Ed} & \text{original} &  $\gamma_l=0.04$ & $\gamma_l=0.08$
& $\gamma_s=0.15$& $\gamma_s=0.05$
                                                  \\
\text{max.~power for}     & & $E_l=7.1$ &  $E_l=12.2$ &  $E_l=3.5$ &
                                                          
                   &  \\
  && $E_s=0.35$&&& $E_s=0.51$&$E_s= 0.14$\\
  \hline
 \text{1 mile (R-pace)} & \text{04:20} & \text{  04:25.21} & \text{  orig.} & \text{  orig.} & \text{  04:16.72} & \text{  04:32.06} \\
 \text{5min} & \text{---} & \text{   04:16.96} & \text{ orig.} & \text{ orig.} & \text{  04:05.87} & \text{  04:27.14} \\
 \text{time $t_c$ (I-pace)} & \text{04:42} & \text{  04:39.22} & \text{ orig.} & \text{ orig.} & \text{ orig.} & \text{  orig.} \\
 \text{5.000m} & \text{  04:49} & \text{  04:49.06} & \text{  04:46.79} & \text{  04:55.54} & \text{   orig.} & \text{   orig.} \\
 \text{10.000m} & \text{  05:00} & \text{  05:00.67} & \text{  04:55.58} & \text{  05:15.85} & \text{   orig.} & \text{   orig.} \\
 \text{60min (T-pace)} & \text{05:06} & \text{  05:03.67} & \text{  04:58.07} & \text{  05:19.64} & \text{   orig.} & \text{   orig.} \\
 \text{Half marathon} & \text{  05:15} & \text{  05:14.30} & \text{  05:05.68} & \text{  05:41.31} & \text{   orig.} & \text{   orig.} \\
 \text{marathon (M-pace)} & \text{05:29} & \text{  05:28.16} & \text{  05:15.70} & \text{  06:09.16} & \text{   orig.} & \text{  orig.} \\     
  \hline
\end{tabular}
\addtolength{\tabcolsep}{4pt}  
\label{tab_vdot40_paces}
\end{adjustwidth}
\end{table}

\begin{table}[H]
\begin{adjustwidth}{-1in}{0in}
\centering
\caption{
{\bf Paces per km for a runner with VDOT=60 score  for different endurances. The original physiological
  parameters are $t_c=12.67$min, $v_m=298.51$m/min, $\gamma_l=0.052$ and
$\gamma_s=0.092$.  The meaning of the columns is the same as in Tab.\ref{tab_vdot40_paces}.}}
\addtolength{\tabcolsep}{-4pt}  
\begin{tabular}{|c|r|r|r|r|r|r|}
  \hline
  \text{pace at} & Ref.~\cite{Daniels3Ed} & \text{original} &  $\gamma_l=0.04$ & $\gamma_l=0.08$
& $\gamma_s=0.15$& $\gamma_s=0.05$
                                                  \\
\text{max.~power for}     & & $E_l=6.8$ &  $E_l=12.2$ &  $E_l=3.5$ &
                                                           
                   &  \\
  && $E_s=0.34$&&&$E_s=0.51$&$E_s= 0.14$\\
  \hline
 \text{1 mile (R-pace)} & \text{03:05} & \text{  03:05.04} & \text{  orig.} & \text{  orig.} & \text{  02:54.93} & \text{  03:12.36} \\
 \text{5min} & \text{---} & \text{  03:05.15} & \text{  orig.} & \text{   orig.} & \text{  02:56.39} & \text{  03:12.07} \\
 \text{time $t_c$ (I-pace)} & \text{03:23} & \text{  03:21.00} & \text{   orig.} & \text{  orig.} & \text{   orig.} & \text{   orig.} \\
 \text{5.000m} & \text{  03:25} & \text{  03:24.14} & \text{  03:23.36} & \text{  03:26.00} & \text{   orig.} & \text{   orig.} \\
 \text{10.000m} & \text{  03:32} & \text{  03:32.41} & \text{  03:29.49} & \text{  03:39.64} & \text{   orig.} & \text{  orig.} \\
 \text{60min (T-pace)} & \text{03:40} & \text{  03:38.78} & \text{  03:34.33} & \text{  03:49.55} & \text{  orig.} & \text{  orig.} \\
 \text{Half marathon} & \text{  03:42} & \text{  03:42.12} & \text{  03:36.53} & \text{  03:56.62} & \text{   orig.} & \text{   orig.} \\
 \text{marathon (M-pace)} & \text{03:52} & \text{  03:51.99} & \text{  03:43.51} & \text{  04:15.08} & \text{   orig.} & \text{   orig.} \\
  \hline
\end{tabular}
\addtolength{\tabcolsep}{4pt}  
\label{tab_vdot60_paces}
\end{adjustwidth}
\end{table}

\begin{table}[H]
\begin{adjustwidth}{-1in}{0in}
\centering
\caption{
{\bf Paces per km for a runner with VDOT=80 score for different endurances. The original physiological
  parameters are $t_c=12.92$min, $v_m=376.85$m/min, $\gamma_l=0.053$ and
$\gamma_s=0.088$. The meaning of the columns is the same as in Tab.\ref{tab_vdot40_paces}.}}
\addtolength{\tabcolsep}{-4pt}  
\begin{tabular}{|c|r|r|r|r|r|r|}
  \hline
  \text{pace at} & Ref.~\cite{Daniels3Ed} & \text{original} &  $\gamma_l=0.04$ & $\gamma_l=0.08$
& $\gamma_s=0.15$& $\gamma_s=0.05$
                                                  \\
\text{max.~power for}     & & $E_l=6.6$ &  $E_l=12.2$ &  $E_l=3.5$ &
                                                          
                   &  \\
  && $E_s=0.32$&&& $E_s=0.51$&$E_s= 0.14$\\
  \hline
 \text{1 mile (R-pace)} & \text{02:25} & \text{  02:23.98} & \text{  orig.} & \text{   orig.} & \text{  02:13.52} & \text{  02:30.46} \\
 \text{5min} & \text{---} & \text{  02:26.99} & \text{   orig.} & \text{  orig.} & \text{  02:19.37} & \text{  02:32.00} \\
 \text{time $t_c$ (I-pace)} & \text{02:41} & \text{  02:39.22} & \text{  orig.} & \text{   orig.} & \text{   orig.} & \text{  orig.} \\
 \text{5.000m} & \text{  02:40} & \text{  02:39.45} & \text{  02:39.39} & \text{  02:39.59} & \text{   orig.} & \text{   orig.} \\
 \text{10.000m} & \text{  02:46} & \text{  02:45.91} & \text{  02:44.14} & \text{  02:49.88} & \text{   orig.} & \text{   orig.} \\
 \text{60min (T-pace)} & \text{02:54} & \text{  02:53.33} & \text{  02:49.64} & \text{  03:01.52} & \text{  orig.} & \text{   orig.} \\
 \text{Half marathon} & \text{  02:53} & \text{  02:53.50} & \text{  02:49.59} & \text{  03:02.65} & \text{  orig.} & \text{  orig.} \\
 \text{marathon (M-pace)} & \text{03:01} & \text{  03:01.22} & \text{  02:54.99} & \text{  03:16.46} & \text{  orig.} & \text{   orig.} \\
  \hline
\end{tabular}
\addtolength{\tabcolsep}{4pt}  
\label{tab_vdot80_paces}
\end{adjustwidth}
\end{table}

\begin{table}[H]
\begin{adjustwidth}{-1in}{0in}
\centering
\caption{
{\bf Paces per km for a runner with VDOT=40 score for different
  variations of the time $t_c$. The original physiological
  parameters are $t_c=12.35$min, $v_m=214.88$m/min, $\gamma_l=0.051$ and
$\gamma_s=0.096$.}}
\addtolength{\tabcolsep}{-4pt}  
\begin{tabular}{|c|r|r|r|r|r|r|}
  \hline
  \text{pace at} & Ref.~\cite{Daniels3Ed} & \text{original} &  $0.8 t_c$ &  $0.9 t_c$
&  $1.1 t_c$ &  $1.2 t_c$
                                                  \\
\text{max.~power for}     & & $t_c=12.35$min &  &&&\\
  \hline
 \text{1 mile (R-pace)} & \text{04:20} & \text{  04:25.21} & \text{  04:31.27} & \text{  04:28.03} & \text{  04:22.70} & \text{  04:20.46} \\
 \text{5min} & \text{---} & \text{  04:16.96} & \text{  04:22.12} & \text{  04:19.37} & \text{  04:14.82} & \text{  04:12.90} \\
 \text{time $t_c$ (I-pace)} & \text{04:42} & \text{  04:39.22} & \text{  04:42.43} & \text{  04:40.73} & \text{  04:36.70} & \text{  04:34.43} \\
 \text{5.000m} & \text{  04:49} & \text{  04:49.06} & \text{  04:52.69} & \text{  04:50.76} & \text{  04:47.53} & \text{  04:46.15} \\
 \text{10.000m} & \text{  05:00} & \text{  05:00.67} & \text{  05:04.62} & \text{  05:02.52} & \text{  04:59.02} & \text{  04:57.52} \\
 \text{60min (T-pace)} & \text{05:06} & \text{  05:03.67} & \text{  05:07.47} & \text{  05:05.45} & \text{  05:02.07} & \text{  05:00.63} \\
 \text{Half marathon} & \text{  05:15} & \text{  05:14.30} & \text{  05:18.63} & \text{  05:16.33} & \text{  05:12.49} & \text{  05:10.86} \\
 \text{marathon (M-pace)} & \text{05:29} & \text{  05:28.16} & \text{  05:32.89} & \text{  05:30.37} & \text{  05:26.18} & \text{  05:24.39} \\
 \hline
\end{tabular}
\addtolength{\tabcolsep}{4pt}  
\label{tab_vdot40_paces_2}
\end{adjustwidth}
\end{table}

\begin{table}[H]
\begin{adjustwidth}{-1in}{0in}
\centering
\caption{
{\bf Paces per km for a runner with VDOT=60 score for different
  variations of the time $t_c$.  The original physiological
  parameters are $t_c=12.67$min, $v_m=298.51$m/min, $\gamma_l=0.052$ and
$\gamma_s=0.092$.}}
\addtolength{\tabcolsep}{-4pt}  
\begin{tabular}{|c|r|r|r|r|r|r|}
  \hline
  \text{pace at} & Ref.~\cite{Daniels3Ed} & \text{original} &  $0.8 t_c$ &  $0.9 t_c$
&  $1.1 t_c$ &  $1.2 t_c$
                                                  \\
  \text{max.~power for}   & & $t_c=12.67$min &  &&&\\
  \hline
 \text{1 mile (R-pace)} & \text{03:05} & \text{  03:05.04} & \text{  03:08.94} & \text{  03:06.86} & \text{  03:03.42} & \text{  03:01.97} \\
 \text{5min} & \text{---} & \text{  03:05.15} & \text{  03:08.72} & \text{  03:06.82} & \text{  03:03.67} & \text{  03:02.34} \\
 \text{time $t_c$ (I-pace)} & \text{03:23} & \text{  03:21.00} & \text{  03:23.37} & \text{  03:22.11} & \text{  03:19.25} & \text{  03:17.68} \\
 \text{5.000m} & \text{  03:25} & \text{  03:24.14} & \text{  03:26.73} & \text{  03:25.36} & \text{  03:23.06} & \text{  03:22.08} \\
 \text{10.000m} & \text{  03:32} & \text{  03:32.41} & \text{  03:35.22} & \text{  03:33.72} & \text{  03:31.23} & \text{  03:30.17} \\
 \text{60min (T-pace)} & \text{03:40} & \text{  03:38.78} & \text{  03:41.60} & \text{  03:40.10} & \text{  03:37.60} & \text{  03:36.54} \\
 \text{Half marathon} & \text{  03:42} & \text{  03:42.12} & \text{  03:45.20} & \text{  03:43.56} & \text{  03:40.83} & \text{  03:39.66} \\
 \text{marathon (M-pace)} & \text{03:52} & \text{  03:51.99} & \text{  03:55.36} & \text{  03:53.57} & \text{  03:50.58} & \text{  03:49.31} \\
 \hline
\end{tabular}
\addtolength{\tabcolsep}{4pt}  
\label{tab_vdot60_paces_2}
\end{adjustwidth}
\end{table}

\begin{table}[H]
\begin{adjustwidth}{-1in}{0in}
\centering
\caption{
{\bf Paces per km for a runner with VDOT=80 score for different
  variations of the time $t_c$. The original physiological
  parameters are $t_c=12.92$min, $\gamma_l=0.053$ and
$\gamma_s0.088$.}}
\addtolength{\tabcolsep}{-4pt}  
\begin{tabular}{|c|r|r|r|r|r|r|}
  \hline
  \text{pace at} & Ref.~\cite{Daniels3Ed} & \text{original} &  $0.8 t_c$ &  $0.9 t_c$
&  $1.1 t_c$ &  $1.2 t_c$
                                                  \\
\text{max.~power for}  & & $t_c=12.92$min &  &&&\\
  \hline
 \text{1 mile (R-pace)} & \text{02:25} & \text{  02:23.98} & \text{  02:26.80} & \text{  02:25.30} & \text{  02:22.81} & \text{  02:21.76} \\
 \text{5min} & \text{---} & \text{  02:26.99} & \text{  02:29.69} & \text{  02:28.25} & \text{  02:25.87} & \text{  02:24.85} \\
 \text{time $t_c$ (I-pace)} & \text{02:41} & \text{  02:39.22} & \text{  02:41.12} & \text{  02:40.11} & \text{  02:37.90} & \text{  02:36.71} \\
 \text{5.000m} & \text{  02:40} & \text{  02:39.45} & \text{  02:41.48} & \text{  02:40.40} & \text{  02:38.17} & \text{  02:36.87} \\
 \text{10.000m} & \text{  02:46} & \text{  02:45.91} & \text{  02:48.11} & \text{  02:46.94} & \text{  02:44.99} & \text{  02:44.16} \\
 \text{60min (T-pace)} & \text{02:54} & \text{  02:53.33} & \text{  02:55.60} & \text{  02:54.39} & \text{  02:52.38} & \text{  02:51.53} \\
 \text{Half marathon} & \text{  02:53} & \text{  02:53.50} & \text{  02:55.91} & \text{  02:54.63} & \text{  02:52.49} & \text{  02:51.58} \\
 \text{marathon (M-pace)} & \text{03:01} & \text{  03:01.22} & \text{  03:03.85} & \text{  03:02.45} & \text{  03:00.11} & \text{  02:59.12} \\
 \hline
\end{tabular}
\addtolength{\tabcolsep}{4pt}  
\label{tab_vdot80_paces_2}
\end{adjustwidth}
\end{table}

\section*{Conclusion}

Modern performance testing is often based on laboratory testing of athletes
with the goal of identifying physiological metrics that correlate with
performance and can be linked to fundamental physiological
processes. However, measuring physiological metrics requires time
consuming and expensive testing, often under rather idealized laboratory
conditions. Hence, it appears to be very useful to extract information on power characteristics for individual runners or certain groups of runners from performance results in racing events or time trails. 
This is of particularly great interest for analyzing the effect of aging on human performance, considering the enormous improvement of performance in older age groups. As stated already by A.~V.~Hill, world and other 
records constitute very interesting data sets since their accuracy by
far exceeds that of laboratory measurements and they correspond to
best human performances at a given time in history under realistic
conditions.

The model presented here provides a quantitative method for extracting  characteristic parameters from  race
performances of a group of runners or of an individual runner.
These parameters quantify the runner's performance status and can be used to predict personalized
fastest possible but realistic and safe racing paces for a wide range of race distances and durations.  Our model provides an unified description of running events at sub- and supra-maximal velocities that are separated by a time scale $t_c$ whose value is in good agreement with independent measurements.
On a fundamental level, for the first time our approach provides a derivation of the previously observed but unexplained linear relation between the mean velocity and the logarithm of the duration for running records. The mechanism underlying this logarithmic relation could be identified as the necessity of a supplemental power, beyond the nominal power cost of running, for maintaining the mean velocity.  Our findings are different from the previously postulated power law relation between the mean race speed $\bar v$
and distance $d$, $\bar v\sim d^{-\beta}$ with an exponent $\beta$
that varies  between $0.054$ and $0.083$, depending
on age and gender \cite{Riegel:1981ci}. Note that this exponent
$\beta$ is slightly smaller than the value $1/8$ expected from
Kennelly's original work \cite{Kennely:1906aa}.  
A modified, broken power law yielded a crossover duration $t_c$ between 3min and 4min which is too short
to be consistent with laboratory measurements \cite{S.:2000fi}. 

We have validated our model by comparing it to various running records and also to personal records of individual runners. The comparison shows consistently low relative errors between actual and predicted race times, with the mean error being maximally $1\%$ and typically less than that for both world and national records and individual personal records. To our knowledge,  this is the to date most accurate theoretical description of running performances that does not require any a priori fixing of physiological constants. The obtain agreement shows that human running performance depends in a subtle manner on several
 variables that, however, can be quantified for
individual runners. Indeed, we find that four parameters can characterize the
performance state of a runner: the time $t_c$ over which the velocity $v_m$ can be maintained, and two endurance parameters 
$E_s$ and $E_l$ for short and long duration endurance. By comparing to independent measurements, we argue that $v_m$ is close to the velocity at maximal aerobic power or VO$_{2max}$. By their definition, the endurance parameters yield the duration $E_l t_c > t_c$ over which a runner can sustain $90\%$ of $v_m$ or maximal aerobic power and the duration $E_s t_c < t_c$ for $110\%$ of $v_m$ or maximal aerobic power.

We have compared our model to Daniel's VDOT model which is based on a single variable parameter (VDOT) that measures performance. When the race times predicted by the VDOT model are analyzed with our model, we find rather superior long distance endurance parameters $E_l$. For more conservative endurance parameters, our model yields marathon race paces that even for elite runners can be 15sec/km slower than the VDOT predictions. This highlights the importance of proper modeling of individual endurances.

Future studies based on our model could include the dependence of the performance state on distance
specialization, altitude, air temperature, age, and other factors.
With the availability of big data set on running performances,
these studies could be performed with much better statistics than studies with much smaller groups of runners participating in laboratory and clinical studies. Our model could be applied to other endurance sports after a modification of the running specific dependence of power on velocity.

\section*{Acknowledgments}

This work was supported by CNRS INP through an EMERGENCE2017 grant and
the A*MIDEX Project ANR-11-IDEX-0001-02 co-funded by the French program
Investissements d’Avenir, managed by the French National Research
Agency. Valuable discussions with Veronique Billat and Francois 
P\'eronnet on various physiological aspects of the model and with Jack Daniels on the methodology of the VDOT model are acknowledged.

\nolinenumbers

%

\section*{S1 Appendix. Solution of the integral equation for $P_{max}(T)$.}

The maximal power $P_{max}(T)$ is determined by the integral equation
\begin{equation}
  \label{eq:S1_inteq}
  P_{max}(T) + P_{sup}(T) = \frac{1}{T} \int_0^T P_{max}(T-t) dt =  \frac{1}{T} \int_0^T P_{max}(t) dt
\end{equation}
with $P_{sup}(T)$ given by Eq.~\eqref{eq:2}. This equation can be easily
transformed into a differential equation by defining the indefinite
integral $E(T)$ of $P_{max}(T)$ so that the derivative  $E'(T) = P_{max}(T)$.
Without loss of generality, we can chose the initial condition
$E(0)=0$. The differential equation for $E(T)$ is then
\begin{equation}
  \label{eq:S1_diff_eq}
  E'(T) +P_{sup}(T) = \frac{E(T)}{T} 
\end{equation}
which has the general solution
\begin{equation}
  \label{eq:S1_diff_solution}
  E(T) = T P_m + T P_{sup}(t_c) - T \int_{t_c}^T \frac{P_{sup}(t)}{t} dt
\end{equation}
where we imposed the initial condition $E'(T=t_c)=P_m$ so that
$P_{max}(T=t_c)=P_m$ as required by definition of $P_m$. Performing the
integral with the constant function $P_{sup}(t)=P_s$ for $T\le t_c$
yields
\begin{equation}
  \label{eq:E_a}
  E(T)=T \left[ P_m + P_s -P_s \log(T/t_c) \right]
\end{equation}
and using $P_{sup}(t) = P_l(t-t_c)/t+P_s t_c/t$ for $T \ge t_c$ yields
\begin{equation}
  \label{eq:E_an}
  E(T)=T \left[ P_m + P_s + (P_l-P_s) \frac{T-t_c}{T} -P_l
    \log(T/t_c) \right] \, .
\end{equation}
Taking the derivative of this solution, we finally obtain the solution
\begin{equation}
  \label{eq:1a}
  P_{max}(T) =   P_m - P_s \log(T/t_c)
\end{equation}
for $T \le t_c$ and
\begin{equation}
  \label{eq:1b}
  P_{max}(T) =  P_m - P_l \log(T/t_c)
\end{equation}
for $T \ge t_c$. This is the result given in Eq.~\eqref{eq:5}.

\section*{S2 Appendix. Comparison to oxygen uptake measurement}

While it is assuring to see below that our model can explain and
predict record and individual racing times, a more direct comparison
to power output during running is desirable to probe the logarithmic
decline of the maximal power output with exercise duration, as
predicted by Eq.~\eqref{eq:5}. This is of particular importance in the
anaerobic range where different functional forms, e.g., exponential
decays, have been proposed \cite{Peronnet:1989dp}.  However, running
power, as measured by oxygen utilization, can be directly determined only in the aerobic regime. For
(supra-maximal) exercise with substantial contributions from anaerobic
systems where power output exceeds maximal oxygen uptake, Medbo et
al. showed that the oxygen demand can be estimated by extrapolating
each runner's individual nominal linear relationship between running speed and
submaximal oxygen uptake \cite{Medbo:ts}. The difference between the extrapolated
oxygen utilization and the measured oxygen uptake is the accumulated
oxygen deficit. Using this method, Medbo et al. determined from
treadmill exercise at speeds that caused exhaustion within different
predetermined durations the oxygen demand relative to the maximal
uptake. Translated to percent of maximal aerobic power output, this
oxygen demand is given by $100 \times P_{max}(T)/P_m$ in our model,
with $P_{max}(T)$ given in Eq.~\eqref{eq:5}.

\begin{figure}[H]
  \centering
 \includegraphics[width=0.8\textwidth]{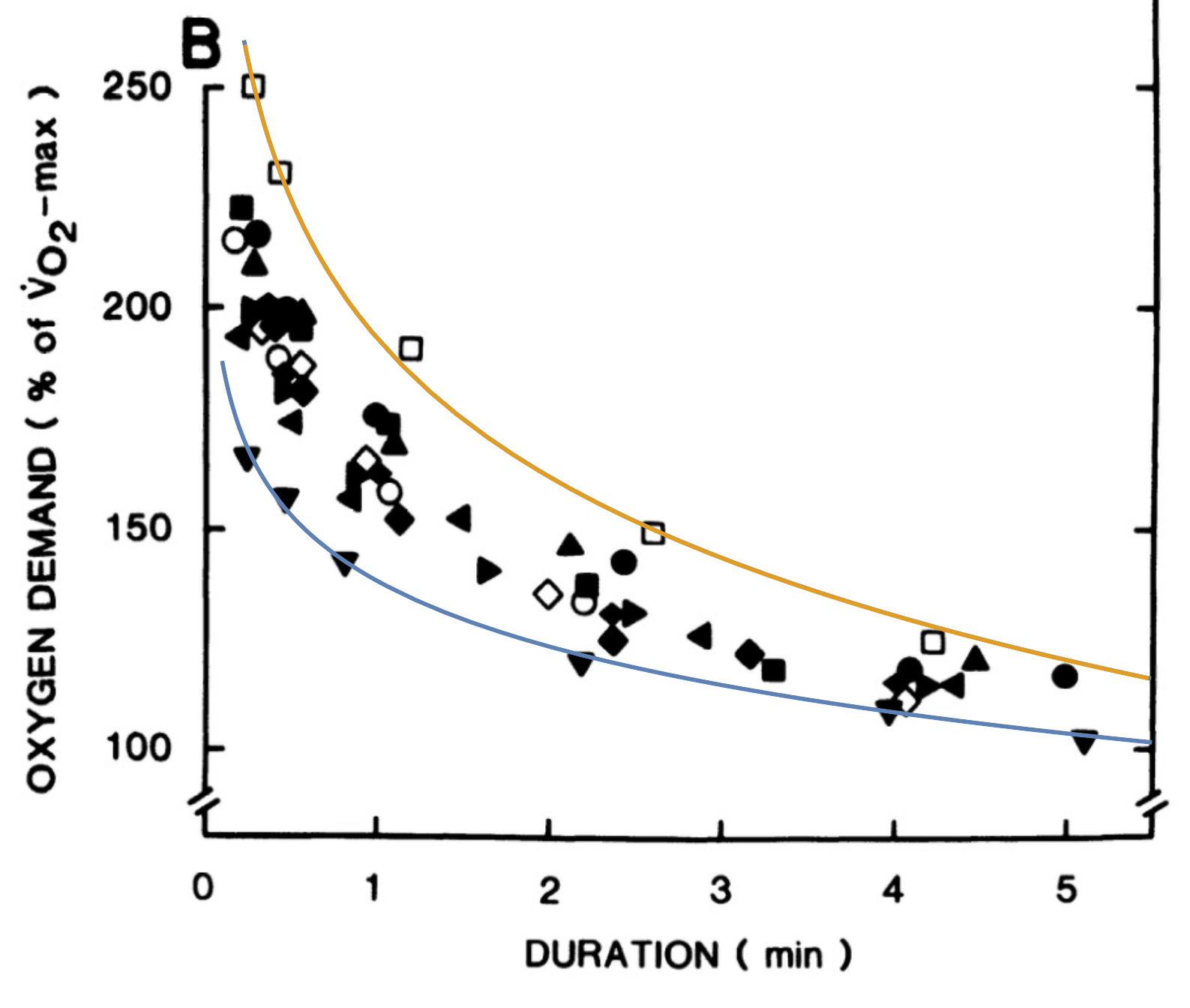}
 \caption{{\bf Relative nominal oxygen demand as function of the
     maximum duration over which it can be sustained. Original plot
     and data for 11 runners from Ref. \cite{Medbo:ts}. The two curves are
     fits of Eq.~\eqref{eq:5} to the data for the runners with
     smallest and largest relative oxygen demand.}}
  \label{fig:VO2_of_Tmax}
\end{figure}

While a logarithmic dependence for $P_{max}(T)$ has been deduced from
purely empirical data analyses for world records for times above
$t_c$ before \cite{Peronnet:1989dp}, to our knowledge a logarithmic
scaling has not been proposed for shorter exercise with large
anaerobic involvement.  Hence, it is interesting that there exists
experimental estimates of the maximal oxygen utilization that can be
maintained for a given duration. As explained above, Medbo et
al. \cite{Medbo:ts} obtained for 11 runners data that correspond to
$100 \times P_{max}(T)/P_m$ which is shown as function of $T < 5$min
$\sim t_c$ in Fig.~\ref{fig:VO2_of_Tmax}. We have fitted the
prediction of our model to the data, and the results for the runner
with smallest and largest oxygen demand are shown in the same
figure. The agreement between the data and our model prediction
appears to be rather convincing. This suggests that there exists
indeed a logarithmic relation between maximally sustainable power and
duration in the range of supra-maximal intensities, resembling
observation that were made before in the sub-maximal zone.

\end{document}